\documentclass{aa}  
\usepackage{graphicx}
\usepackage{amsmath}
\usepackage{amssymb}
\usepackage{gensymb}
\usepackage{natbib}
\usepackage{booktabs}
\usepackage{multirow}
\usepackage{pdflscape}
\usepackage{float}
\usepackage{subfig}
\usepackage{color}
\usepackage{xcolor}
\usepackage[font=small]{caption}
\usepackage{mathtools, cuted}
\usepackage[normalem]{ulem}
\usepackage[labelfont=bf]{caption}

\title{Ultra-high-energy cosmic rays and neutrinos \\ from tidal disruptions by massive black holes}
               
\authorrunning{C. Gu\'epin et al.}
\titlerunning{Ultra-high-energy cosmic rays and neutrinos from tidal disruptions}

\author{Claire Gu\'epin\inst{1,2,3}, Kumiko Kotera\inst{1}, Enrico Barausse\inst{1,4,5}, Ke Fang\inst{6} and Kohta Murase\inst{7,8,9,10}}

\institute{Sorbonne Universit\'e, CNRS, UMR 7095, Institut d’Astrophysique de Paris, 98 bis bd Arago, 75014 Paris, France
\and Department of Astronomy, University of Maryland, College Park, MD  20742, USA
\and Joint Space-Science Institute, University of Maryland, College Park, MD  20742, USA
\and SISSA, Via Bonomea 265, 34136 Trieste, Italy and INFN Sezione di Trieste
\and Institute for Fundamental Physics of the Universe, Via Beirut 2, 34014 Trieste, Italy
\and Kavli Institute for Particle Astrophysics and Cosmology, Stanford University, Stanford, CA 94305, USA
\and Department of Physics, The Pennsylvania State University, University Park, PA 16802, USA
\and Department of Astronomy \& Astrophysics, The Pennsylvania State University, University Park, PA 16802, USA
\and Center for Particle and Gravitational Astrophysics, The Pennsylvania State University, University Park, PA 16802, USA
\and Yukawa Institute for Theoretical Physics, Kyoto, Kyoto 606-8502 Japan}

\date{ }

\abstract{Tidal disruptions are extremely powerful phenomena that have been designated as candidate sources of ultra-high-energy cosmic rays. The disruption of a star by a black hole can naturally provide protons and  heavier nuclei, which can be injected and accelerated to ultra-high energies within a jet. Inside the jet, accelerated nuclei are likely to interact with a dense photon field, leading to a significant production of neutrinos and secondary particles. We model numerically the propagation and interactions of high-energy nuclei in jetted tidal disruption events in order to evaluate consistently their signatures in cosmic rays and neutrinos. We propose a simple model of the  light curve of tidal disruption events,  consisting of two stages: a {high state} with bright luminosity and short duration and a {medium state}, less bright and longer lasting. These two states have different impacts on the production of cosmic rays and neutrinos. In order to calculate the diffuse fluxes of cosmic rays and neutrinos, we model the luminosity function and redshift evolution of jetted tidal disruption events. We find that we can fit the latest ultra-high-energy cosmic-ray spectrum and composition results of the Auger experiment for a range of reasonable parameters. The diffuse neutrino flux associated with this scenario is found to be subdominant, but nearby events can be detected by IceCube or next-generation detectors such as IceCube-Gen2.\vspace{1cm}}

\begin{document}

\renewcommand\d{\mathrm{d}}

\maketitle

\section{Introduction}

The detection of ultra-high-energy cosmic rays (UHECRs) with energies $> 10^{20}\,{\rm eV}$ implies the existence of extremely powerful astrophysical accelerators that have not yet  been identified. Several UHECR source models have been proposed in the literature, such as radio-loud active galactic nuclei (AGN),  cluster accretion shocks for steady objects, gamma-ray bursts (GRBs), fast-rotating neutron stars, or giant AGN flares for transient candidates  (see, e.g., \citealp{Kotera11} and references therein). Most of these models can successfully fit the observational data of the Auger and Telescope Array experiments for specific choices of astrophysical parameters, and predict associated high-energy neutrino fluxes that could be observed in the next decade by existing and future experiments  (see, e.g., \citealp{Guepin17} for a systematic study of neutrino signals from transient sources). With the current set of data, however, there is  no evidence that allows us to strongly favor one particular scenario over the others. 

Many other types of transient powerful events have been discovered lately thanks to unprecedented instrumental performance in terms of time resolution and sensitivity. Among them, tidal disruption events (TDE), and more specifically jetted TDEs  observed by the {\it Swift} detector \citep[e.g.,][]{Cummings11,Cenko12}, appear to be interesting candidate sources of  UHECRs, with their impressive energy reservoirs and estimated occurrence rates.  

Tidal disruption events can occur when stars approach massive black holes located at the center of galaxies at distances smaller than the tidal disruption radius. If this radius  is larger than the Schwarzschild radius of the black hole, tidal forces can violently disrupt the star and produce luminous and long-lasting flares. After the disruption of the stellar object, which might be a main sequence star or in some extreme cases a white dwarf,  part of its material escapes and part is accreted, launching simultaneously a wind or a relativistic outflow. 

These transient events were predicted theoretically about 20 years before their first detections, and TDEs lasting for months (or sometimes years) have been observed in the UV, X-rays and $\gamma$ rays  \citep[e.g.,][]{Komossa15}. The emission mostly shows a fast rising phase and a luminosity decay $L \propto t^{-5/3}$, coherent with fallback accretion \citep{Phinney89}. The most luminous events show a higher variability, with sequences of flares of $\sim 1000\,$s alternating with quiescent periods of $\sim 5 \times 10^4\,$s. As they can reach luminosities of $L_{\rm max} = 10^{48}\,{\rm erg\,s}^{-1}$, and can maintain very high bolometric luminosities ($L_{\rm bol} \sim 10^{47}\,{\rm erg\,s}^{-1}$) lasting about $10^6\,{\rm s}$, these powerful emissions are very likely to come from a relativistic jet launched from the central massive black hole \cite[e.g.,][]{Bloom2011, Burrows2011}. To date, it is still not clear if non-jetted and jetted TDEs constitute two distinct populations.

Jetted TDEs could be an ideal site for the production of UHECRs (via the injection of the disrupted stellar material and its acceleration in the jet) and for the production of high-energy neutrinos (produced later by the interaction of the accelerated hadrons with the ambient radiative and/or hadronic backgrounds). Although only a handful of jetted TDEs have been detected so far, these objects have already attracted great interest in the high-energy astroparticle community. High-energy neutrino production in TDE jets was considered before the discovery of IceCube neutrinos \citep[e.g.,][]{Murase08a,MT09,Wang11}, and contributions to IceCube neutrinos have been studied~\citep{Senno16b,Dai16,Lunardini16,Wang16}.  

The UHECR production in TDE jets was originally suggested by \cite{Farrar09}, and the external shock scenario was also considered in detail \citep{farrar2014}. However, it should be kept in mind that the rate of TDEs is too tight to fit the observed UHECR fluxes, as can be deduced from the constraints derived by  \cite{Murase09}, who obtained $\dot{n}_{\rm tde}>1~{\rm Gpc}^{-3}~{\rm yr}^{-1}$. Hence a pure proton case is disfavored and the nucleus scenario is required.
Recent studies attempted to inject a mixed composition and fit the UHECR flux and composition simultaneously in {both}  the internal and external shock scenarios \citep{AlvesBatista17,Zhang17}.

So far, the existing studies have not attempted to model the production of UHECRs in the inner part of the TDE jet (with acceleration occurring at internal shocks for instance). Modeling this effect requires taking into account the interaction of accelerated nuclei inside the jet in order to calculate consistently the resulting chemical composition. In this work, we  study the interaction of accelerated nuclei inside the TDE jet, and the signatures they can produce in UHECRs and neutrinos. For this purpose we  developed a new propagation and interaction code that is comprised of modules from CRPropa3 \citep{AlvesBatista16} and from the code described in \cite{Kotera09}.

In order to calculate the diffuse fluxes of UHECRs and neutrinos, we also introduce a new model for the event rate evolution and luminosity function of TDEs powering jets. The semi-analytic galaxy formation model of \cite{Barausse2012} is used to model the cosmological evolution of massive black holes, which can be related to the jetted TDE comoving event rate density, and thus to the diffuse UHECR and neutrino fluxes.

The properties of TDEs powering jets are subject to large uncertainties. Therefore, we scan the parameter space allowed by TDE observations to model the radiation region (Sect.~\ref{Sec::model}) and the typical photon field inside a TDE jet (Sec.~\ref{Sec::SED}). Inside this region, we consider different interaction processes, detailed in Sect.~\ref{Sec::processes_p}. We then calculate mean free path (MFP) tables for the interaction of protons and heavier nuclei with the photon field of the jet. We use these tables in our code to predict UHECR and neutrino signatures (Sect.~\ref{Sec::results}) for single sources. In order to estimate the diffuse particle fluxes from a population of jetted TDEs, we derive the luminosity function and occurrence rate evolution of these events (Sect.~\ref{section:population}). We find in Sect.~\ref{section:diffuse} that we can fit the latest UHECR spectrum and composition results of the Auger experiment for a range of reasonable parameters. The diffuse neutrino flux associated with this scenario is found to be detectable with IceCube in the next decade. Transient neutrino signals from single sources would be difficult to detect by IceCube or the upcoming GRAND experiment, except for sources located within $\sim 50$~Mpc, associated with a very low event rate.

\section{Interaction of UHE nuclei inside TDE jets}

In the following, all primed quantities are in the comoving frame of the emitting region. Other quantities are in the observer frame. Quantities are labeled $Q_x\equiv Q/10^x$ in cgs units unless specified otherwise, and except for particle energies, which are in $E_x\equiv E/10^x\,$eV.

The tidal disruption of a stellar object can occur if it gets close enough to a black hole, and will produce observable flares if it happens outside the black hole event horizon. A part of the stellar material forms a thick accretion disk, and a fraction of this material  accretes onto the black hole,  most likely in a super-Eddington regime. For most TDEs, the observed radiation comes from the dissipation inside the accretion disk, characterized by a thermal spectrum peaking in extreme ultraviolet or soft X-rays; for a rotating black hole launching a relativistic jet, a non-thermal hard X-ray radiation can be detected, presumably produced through synchro-Compton radiation \citep[e.g.,][]{Burrows2011}. The jet radiation should dominate the observed spectrum for black holes with low mass and high spin, jets oriented toward the observer, and large radiative efficiency of the jet.

In the jet comoving frame, using a condition of causality, $R' \simeq \Gamma c \,t_{\rm var}$ can be considered as the size of the emitting region. In the internal shock model, the distance of the emission region from the black hole is estimated to be $R \simeq \Gamma^2 c \,t_{\rm var} = 3\times 10^{14} \,{\rm cm} \,\Gamma_{1}^2 \,t_{{\rm var},2}$, where $t_{\rm var}=10^2\,t_{\rm var,2}$s and $\Gamma=10\Gamma_1$ are the typical variability timescale and bulk Lorentz factor for jetted TDEs, respectively. This radius coincides with the radius estimated from high-latitude emission with a duration of $\sim t_{\rm var}$ (e.g., \citealp{Piran04}). We note that more generally the relationship between $R$ and $R'$ can be modified, for example by  subsequent internal shocks caused by merged shells and the existence of multiple emission regions such as subjets \citep[e.g.,][]{Murase06,Bustamante17}. However, as long as we consider internal shocks in the jet that expand conically, it is reasonable to consider the expression of $R$ obtained for the one-zone calculation, as has been done in the literature of  GRBs.

First, we assume that cosmic rays are injected at the center of a non-evolving radiation region in the comoving frame. The evolution of the region would mainly result in the dilution over time of the radiation and magnetic energy densities, together with adiabatic losses, associated with observable spectral changes. We account for these effects, to a first approximation, by considering two dominant stages for our TDEs: the early stage, when the source is in a {high state}, at its maximum brightness; and a medium state, reached later, for which the source is typically $1-1.5$ orders of magnitude less luminous, but for a longer integrated time. We  argue in the following that these two states have different impacts on the production of UHECRs and their associated neutrinos.

\subsection{UHECR injection and energetics}\label{Sec::model}
Cosmic  ray nuclei from the stellar material can be accelerated to ultra-high energies inside the TDE jet via one of the various mechanisms advocated for GRBs or AGN jets. We assume that acceleration leads to a rigidity-dependent spectrum in 
${\rm d}N_{\rm CR}/{\rm d}E' =  {\cal A} \sum_Z f_Z {E'}^{-\alpha} \exp(-E'/E'_{Z, {\rm max}})$ with an exponential cutoff at $E'_{\rm Z,max}$ for nuclei of charge $Z$. Here ${\cal A}$ is a normalization constant and $f_Z$ is the fraction of elements with charge number $Z$, such that $\sum_Z f_Z = 1$. The spectral index $\alpha$ can vary (typically between $\alpha\sim1$ and $\alpha\gtrsim 2$) depending on the acceleration mechanism (e.g., magnetic reconnection or diffusive shock acceleration). The cosmic-ray composition depends on the composition of the disrupted object, but it also strongly depends on what happens to the elements before they get injected and accelerated in the jet. Heavy nuclei could indeed undergo fragmentation during the disruption of the stellar object, or a large fraction of light nuclei could escape as part of the expelled stellar envelope. In this work, the elements injected in the radiation region are protons (p), helium (He), carbon and oxygen (CO), silicium (Si), and iron (Fe).

The maximum injection energy $E'_{Z, \rm max}$ is determined by the competition between the acceleration timescale for a nucleus of charge $Z$, $t'_{\rm acc} = \eta_{\rm acc}^{-1} E'/ c\,Z\,e\,B'$\footnote{Some papers adopt $t_{\rm acc}=\eta_{\rm acc}\, r'_L/c$ due to the historical convention.}, and the energy loss timescales $t'_{\rm loss}= \min(t'_{\rm dyn},t'_{\rm syn}, t'_{\rm IC}, t'_{\rm BH}, t'_{p\gamma},...)$, where $t'_{\rm dyn}=R / \beta \Gamma c$ is the dynamical timescale (see Appendix~\ref{Appendix::Emax} for numerical estimates of $E'_{Z, \rm max}$). The factor $\eta_{\rm acc} \leq 1$ describes the efficiency of the acceleration process; for a maximally efficient acceleration, $\eta_{\rm acc} = 1$. In this study we neglect the re-acceleration of secondary particles, and leave it for future work.

From the energetics point of view, the luminosity injected into cosmic rays is considered  related to the bolometric luminosity in photons, such that $L_{\rm CR} = \xi_{\rm CR} L_{\rm bol}$,  where we define the baryon loading fraction $\xi_{\rm CR}$ as the fraction of the bolometric luminosity that is injected into cosmic rays of energy $E\ge E_{\rm min}\equiv10^{16}\,$eV.

\subsection{Modeling the TDE spectral energy distribution}\label{Sec::SED}

As suggested in \cite{Senno16b}, we model the spectral energy distribution (SED) inside the TDE jet as a log-parabola with three free parameters: the peak luminosity $L_{\rm pk}$, peak energy $\epsilon_{\rm pk}$, and width $\hat{a}$. The photon energy density then reads
\begin{equation}\label{eq:logparabola}
\epsilon'^2 n_{\epsilon'}' =  \frac{L_{\rm pk}}{4\pi \Gamma^4 R'^2 c}(\epsilon'/\epsilon'_{\rm pk})^{-\hat{a}\log(\epsilon'/\epsilon'_{\rm pk})}\,.
\end{equation}
The peak luminosity and peak energy set the maximum of the SED. The data can help to constrain the width of the log-parabola and a potential high-energy cutoff. However, there are large uncertainties on the observed photon density, due to galaxy absorption, and even more on the photon density inside the jet (see \citealp{Burrows2011,Bloom2011} for the spectrum of Swift J1644+57). 

From our SED model, the bolometric luminosity can then be defined as the luminosity integrated over the entire spectrum: $L_{\rm bol} = \int {\rm d}\epsilon' \, L_{\rm pk}/\epsilon'  \,(\epsilon'/\epsilon'_{\rm pk})^{-\hat{a}\log(\epsilon'/\epsilon'_{\rm pk})}$. As we consider a constant photon field, this bolometric luminosity is larger than the peak luminosity. Moreover, as we model the radiation field inside the jet, we should have $L_{\rm jet, obs} \sim L_{\rm bol}$. We note that in most cases, the main contribution to the observed luminosity is the jet luminosity, but for high black hole masses ($M_{\rm bh} > 5 \times 10^7\,M_\odot$), the thermal luminosity is of the same order of magnitude as the jet luminosity \citep{Krolik12}.

\begin{figure}[!tp]
\centering
\includegraphics[width=0.49\textwidth]{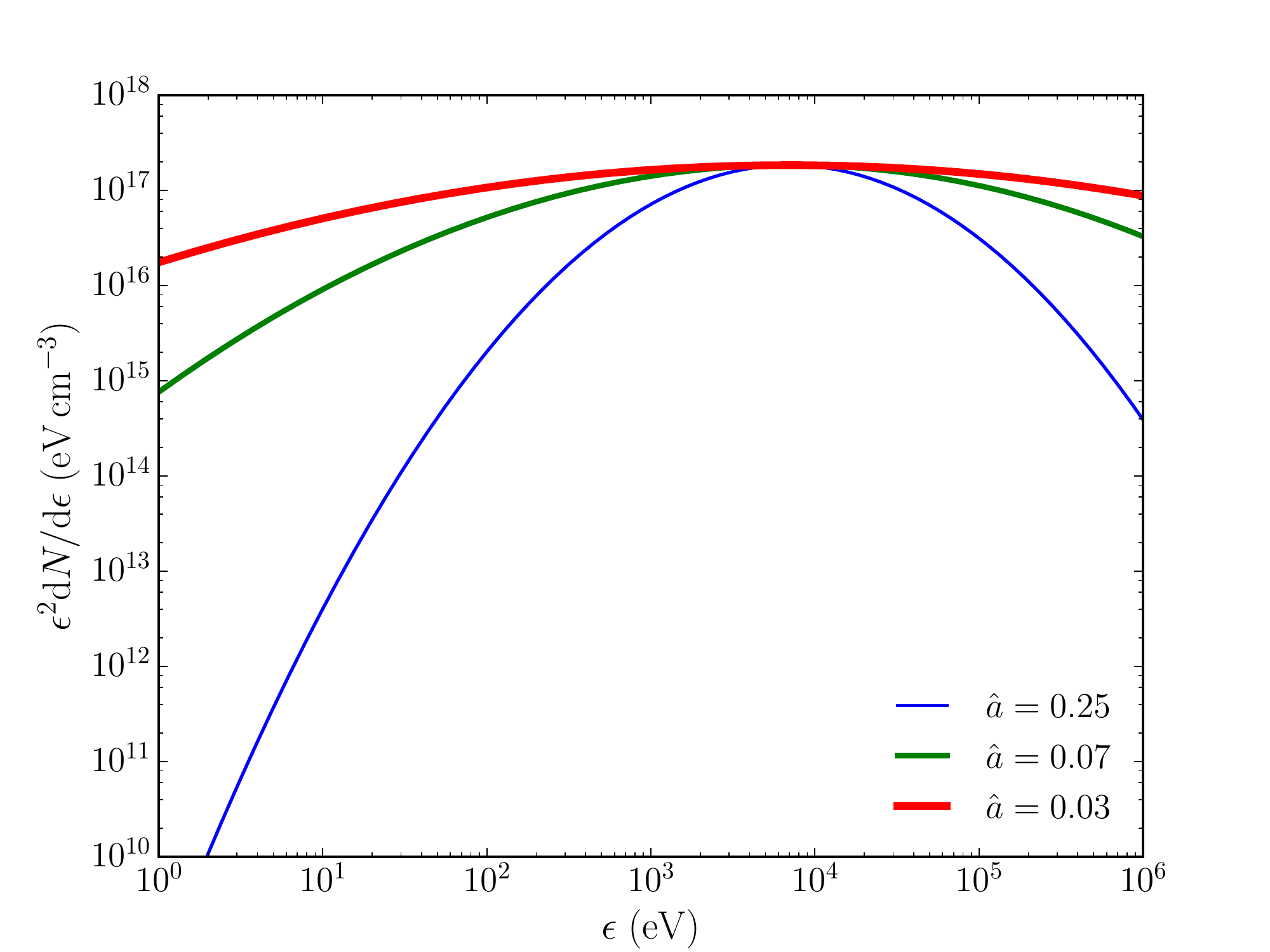}
\caption{ Photon density for a log-parabola model with fixed peak luminosity $L_{\rm pk}=10^{48}\,{\rm erg \,s}^{-1}$.
}\label{Fig::WidthL_comp}
\end{figure}

In this work, we examine several cases summarized in Table~\ref{Tab::tests} and illustrated in Fig.~\ref{Fig::WidthL_comp}. We choose to only vary the width $\hat{a}$ and the peak luminosity $L_{\rm pk}$ of the log-parabola, and to consider a typical peak energy $\epsilon_{\rm pk} = 70\,{\rm keV}$, which is compatible with Swift J1644+57 observations \citep[e.g.,][]{Burrows2011}. Each case corresponds to a different magnetic field, and therefore corresponds to a maximum proton energy $E'_{p,{\rm max}}$ (Eq.~\ref{Eq:Epmax}). The magnetic field is inferred assuming equipartition between the radiative and magnetic energy densities: $\xi_B \int {\rm d}\epsilon' \epsilon' n'_{\epsilon'} = B^2/{8\pi}$ with $\xi_B=1$. Rough equipartition is a standard hypothesis for jets that can be argued from measurements of the energy repartition in extragalactic objects, for example  blazar jets \citep{Celotti08}. It also naturally arises if relativistic reconnection is at play in the outflow, dissipating electromagnetic energy into kinetic energy \citep{Sironi15}.

\begin{table}[t]
\begin{center}
\caption{Properties of the different TDE photon fields considered in this work for a cosmic-ray acceleration efficiency $\eta_{\rm acc}=1$.}\label{Tab::tests}
\begin{tabular}{ccccc}
\toprule  $L_{\rm bol} [L_{\rm pk}]$\,(erg\,s$^{-1}$)  &$\hat{a}$ & $B'\,({\rm G})$ & $E'_{p,{\rm max}} \, ({\rm eV})$ \\
\midrule
\midrule  $3.5\times 10^{48}\,[10^{48}]$  &  $0.25$ & $5.1\times10^3$ & $1.8 \times 10^{18}$ \\
\midrule  $6.8\times 10^{48}\,[10^{48}]$  &  $0.07$ & $7.0\times10^3$ & $2.4 \times 10^{18}$ \\
\midrule  $1.0\times 10^{49}\,[10^{48}]$  &  $0.03$ & $8.7\times10^3$ & $2.2 \times 10^{18}$ \\
\midrule
\midrule  $6.8\times 10^{46}\,[10^{46}]$ &  $0.07$ & $7.0\times10^2$ & $6.3 \times 10^{18}$ \\
\midrule  $6.8\times 10^{47}\,[10^{47}]$ &  $0.07$ & $2.2\times10^3$ & $4.3 \times 10^{18}$ \\
\midrule  $6.8\times 10^{48}\,[10^{48}]$ &  $0.07$ & $7.0\times10^3$ & $2.4 \times 10^{18}$ \\
\bottomrule
\end{tabular}
\vspace{0.3cm}

{\raggedright \small {\bf Notes.} All the photon fields are modeled by a log-parabola (Eq.~\ref{eq:logparabola}), with bolometric luminosity $L_{\rm bol}$, peak luminosity $L_{\rm pk}$, peak energy $\epsilon_{\rm pk}$, and width $\hat{a}$.\par}
\end{center}
\end{table}

The TDE photon spectra evolve in time. As mentioned earlier, although we do not account for proper time evolutions of the SED in this paper, we  consider two states of the SED, inferred from the observations of Swift J1644+57, and which are important for our framework: an early state, corresponding to a high state that can  typically last $t_{\rm dur}\sim 10^5\,$s with a bright luminosity, a high jet efficiency, and a narrow jet opening angle; and a medium state,  $1-1.5$ orders of magnitude less bright, but lasting $t_{\rm dur}\sim 10^6\,$s, with a lower jet efficiency and a similar jet opening angle. For both states, we set a width $\hat{a}=0.07$. These parameters are overall compatible with Swift J1644+57 SED models corrected for galactic absorption \citep[e.g.,][]{Burrows2011}.

\begin{figure*}[!tp]
\centering
\includegraphics[width=\textwidth]{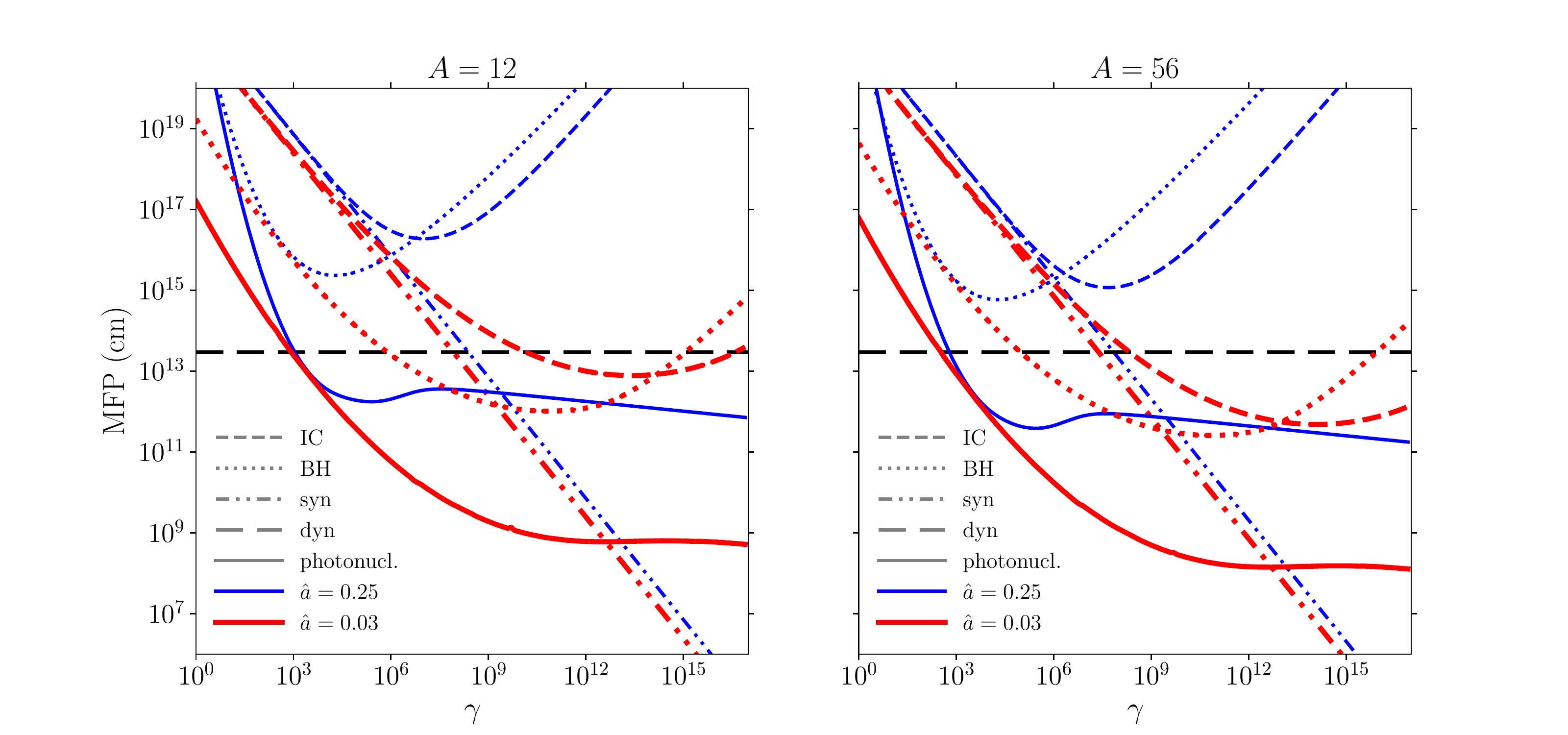}
\caption{Influence of the log-parabola width $\hat{a}$ on the MFPs and energy loss lengths in the comoving frame for carbon (left panel) and iron (right panel) nuclei of Lorentz factor $\gamma$. The peak luminosity is set to $L_{\rm pk}=10^{48}\,{\rm erg \, s}^{-1}$ and two examples of widths are presented: $\hat{a}=0.25$ (blue) and $\hat{a}=0.03$ (red). The different line styles correspond to different processes: photonuclear (solid), inverse Compton (dashed), Bethe--Heitler (dotted), and synchrotron (double dot-dashed). The black long-dashed line corresponds to the typical comoving size of the region. Wider SED lead to larger MFPs.}\label{Fig::MFPN_comp}
\end{figure*}

\begin{figure*}[!tp]
\centering
\includegraphics[width=\textwidth]{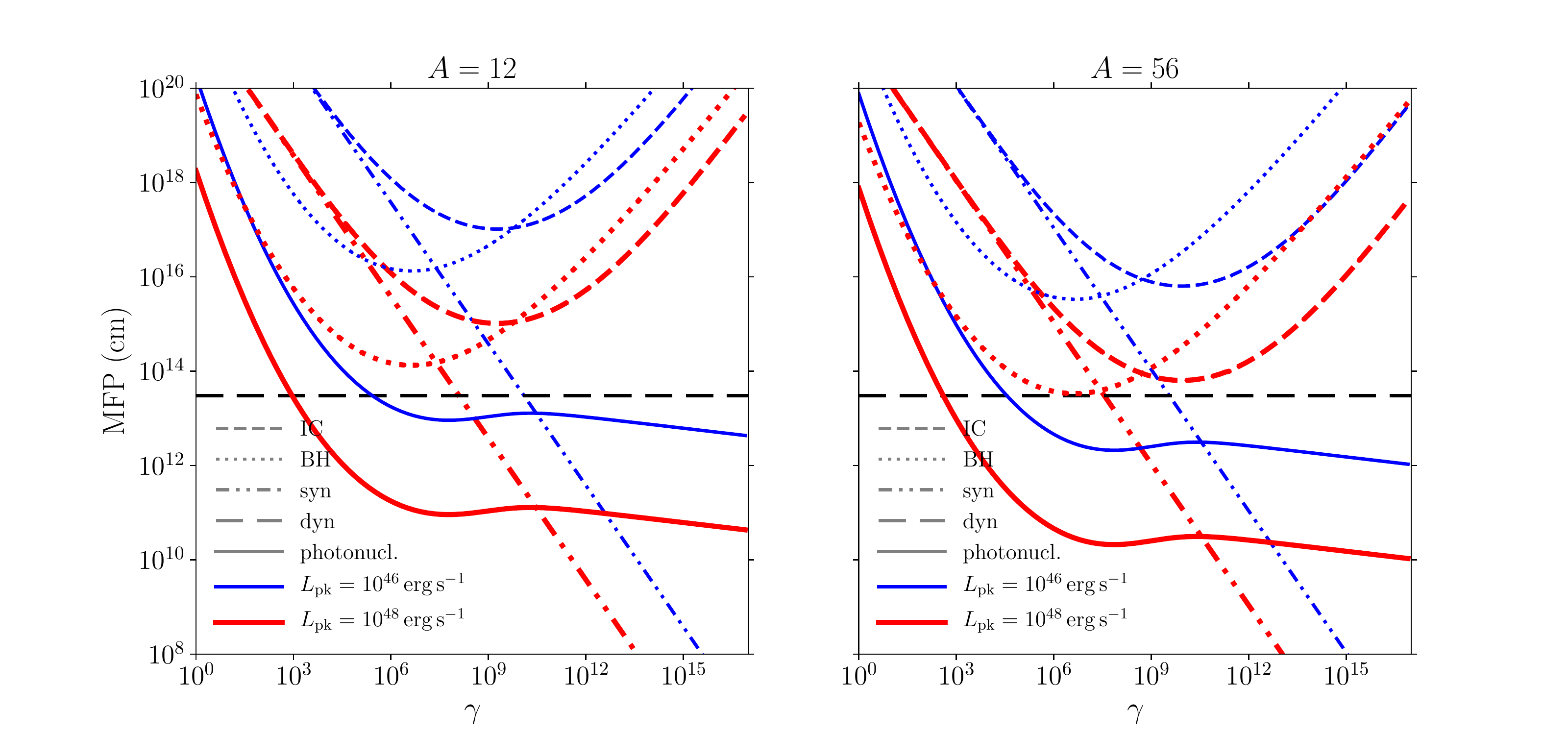}
\caption{Influence of the log-parabola peak luminosity $L_{\rm pk}$. Same as Fig.~\ref{Fig::MFPN_comp}, but for fixed width $\hat{a}=0.07$ and varying the peak luminosity: $L_{\rm pk}=10^{46}\,{\rm erg \, s}^{-1}$ (blue) and $L_{\rm pk}=10^{48}\,{\rm erg \, s}^{-1}$ (red). Higher peak luminosity leads to shorter MFPs.}\label{Fig::MFPN_L_comp}
\end{figure*}

\subsection{Interaction processes}\label{Sec::processes_p}
All relevant interaction processes for nucleons and heavier nuclei are taken into account in our calculations. Nucleons experience pion production via photohadronic and hadronic interactions, as well as neutron decay. Nuclei undergo photonuclear processes in different regimes (requiring increasing photon energy in the nucleus rest frame): giant dipole resonance, quasi deuteron, baryon resonance, and photofragmentation. For all particles, including secondary pions and muons, we account for the synchrotron, inverse Compton, and Bethe--Heitler processes.

All the interaction cross sections and products are obtained from analytic formulae \citep[e.g.,][]{R96, Dermer09} or tabulated using numerical codes like \textsc{Sophia} \citep{Mucke00} for photopion production, \textsc{Talys} \citep{Talys} for photonuclear interactions, and \textsc{Epos} \citep{Werner06} for purely hadronic interactions. We assume that the photofragmentation products are similar to the products of hadronic interactions, which is reasonable to a first approximation. In principle, \textsc{Epos} generates too many free nucleons in the fragmentation process. The scaling between photofragmentation and hadronic interactions is given as a function of the center of mass energy. However, the discrepancy is less than a factor of 2 compared to the data. We note in any case that we  consider here an energy range where uncertainties are very large, and that the  photofragmentation process for nuclei is not dominant in our study.

For extragalactic propagation, photonuclear cross sections and EBL models have a strong influence on the spectrum and composition of cosmic rays, as discussed in \cite{Batista15}. For different EBL models, the discrepancy between cosmic-ray spectra can reach $\sim 40 \%$. The impact of photonuclear models is also strong, whereas more difficult to quantify, especially regarding the channels involving $\alpha$-particles.
 
We show in Figure~\ref{Fig::MFPN_comp} the MFPs or energy loss lengths derived for different photon fields.
In the top panel, we compare the specific luminosities and photon densities for $\hat{a}=0.25$, $\hat{a}=0.07$, and $\hat{a}=0.03$ for a fixed $L_{\rm pk}=10^{48}\,{\rm erg \,s}^{-1}$. The width of the log-parabola has a strong influence on the MFPs as it substantially changes the radiation energy density. The MFPs for the carbon and iron cases are shown in Figure~\ref{Fig::MFPN_comp} for the extreme cases $\hat{a}=0.25$ and $\hat{a}=0.03$. We see that overall, photonuclear interactions dominate over a wide range of particle Lorentz factors $\gamma$, up to ultra-high energies where synchrotron losses start taking over. Changing the width of the log-parabola modifies the MFPs by several orders of magnitude, with shorter paths for narrower SED.

In the bottom panel, we compare the specific luminosities and photon densities for $L_{\rm pk}=10^{46}\,{\rm erg \, s}^{-1}$ and $L_{\rm pk}=10^{48}\,{\rm erg \, s}^{-1}$ for a given width $\hat{a}=0.07$. The influence of peak luminosity on the MFPs is more moderate; as expected, the MFPs are a power of the peak luminosity, and a higher $L_{\rm pk}$ leads to shorter MFP.

\section{UHECRs and neutrinos from single TDEs}\label{Sec::results}

We calculate the cosmic-ray and neutrino spectra after the propagation of protons or nuclei through the photon field of a jetted TDE. The production of neutrinos should be dominated by the high state when the photon field is brightest (and the opacities greatest), and the UHECR production should be calculated over the longer medium state with a lower luminosity but over longer production timescales. We thus calculate the neutrino and UHECR fluxes at their maximum production states. We first consider one single source and show the outgoing spectra for cosmic rays and neutrinos.

\begin{figure*}[!h]
\centering
\includegraphics[width=0.49\textwidth]{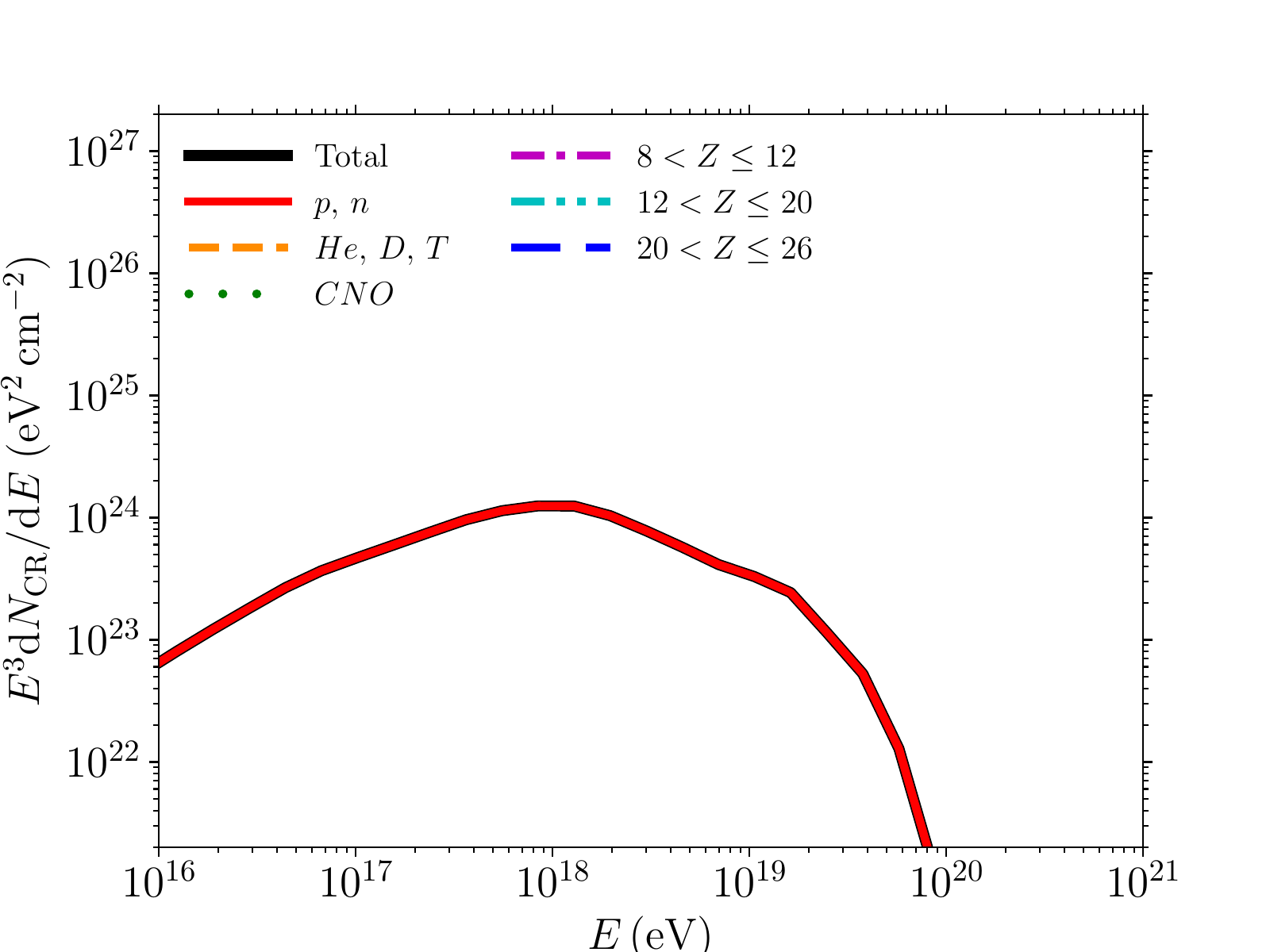}
\includegraphics[width=0.49\textwidth]{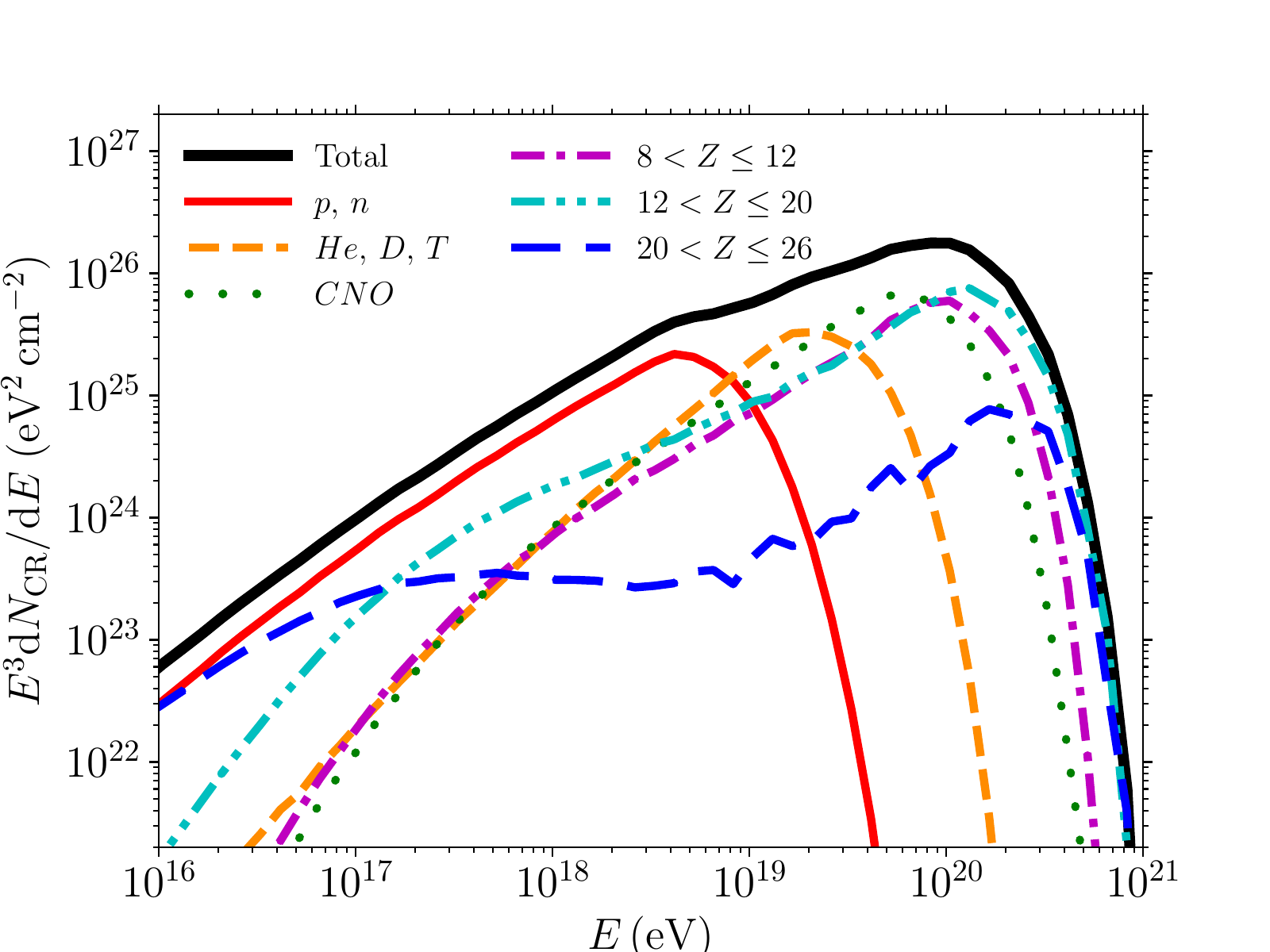}
\caption{Cosmic-ray spectra for one source with pure iron injection with spectral index $\alpha=1.8$, photon field with $\hat{a}=0.07$, and acceleration efficiency $\eta_{\rm acc}=0.2$. We show the total spectrum (black) and the composition (other colors), as indicated in the legend, for TDE around a black hole of mass $M_{\rm bh} = 7 \times 10^{6} M_\odot$, with a corresponding SED in its high state with $L_{\rm pk}=10^{47.5}\,{\rm erg\,s}^{-1}$ and $t_{\rm dur}=10^5\,{\rm s}$ (left) and in its medium state with $L_{\rm pk}=10^{46}\,{\rm erg\,s}^{-1}$ and $t_{\rm dur}=10^6\,{\rm s}$ (right) for a source distance $d_{L,1} =1.88\,{\rm Gpc}$. We assume here $\xi_{\rm CR}=1$.}\label{Fig:CR_Fe_1e46_007_110_10}
\end{figure*}

\begin{figure*}[!h]
\centering
\includegraphics[width=0.49\textwidth]{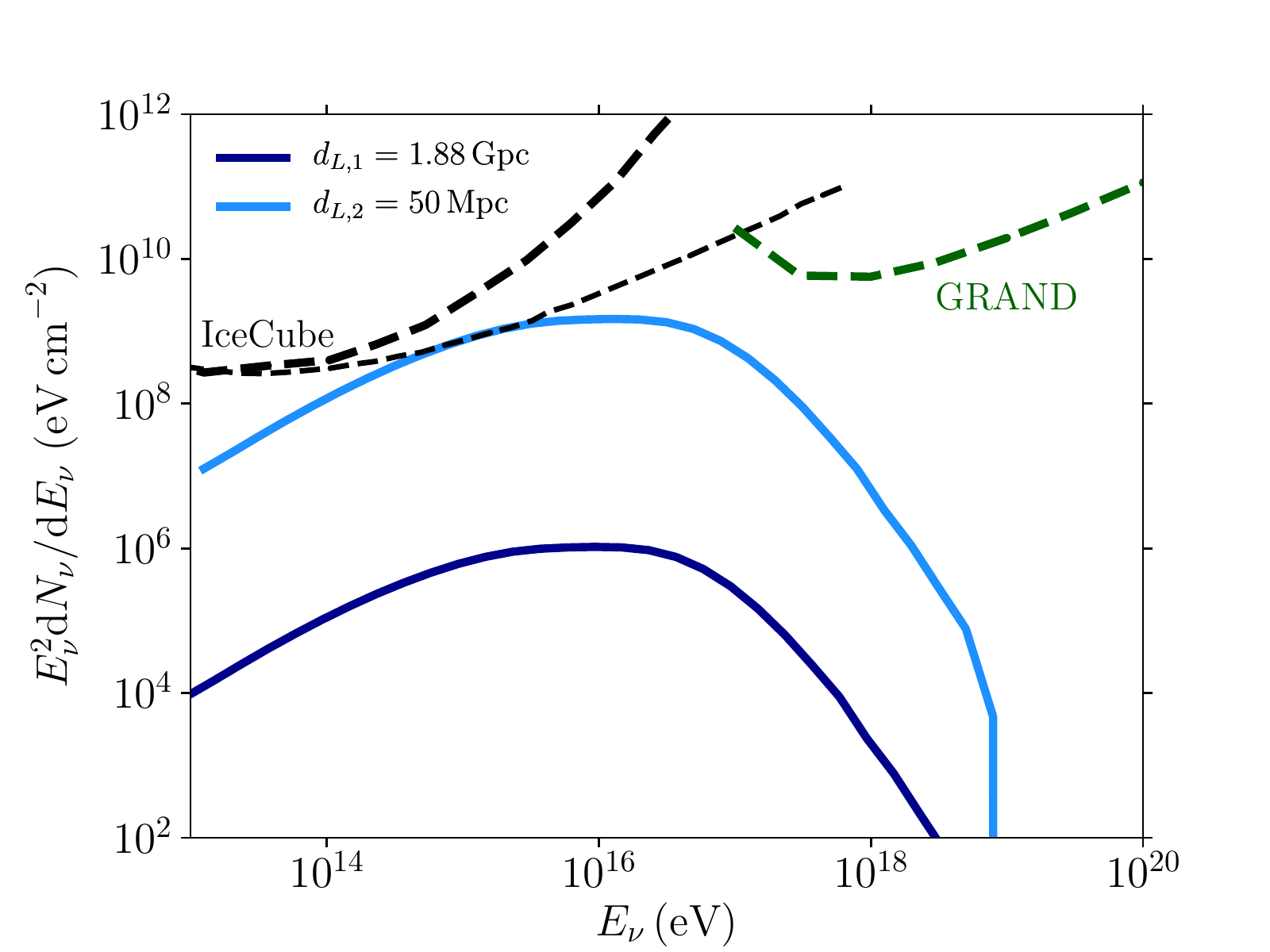}
\includegraphics[width=0.49\textwidth]{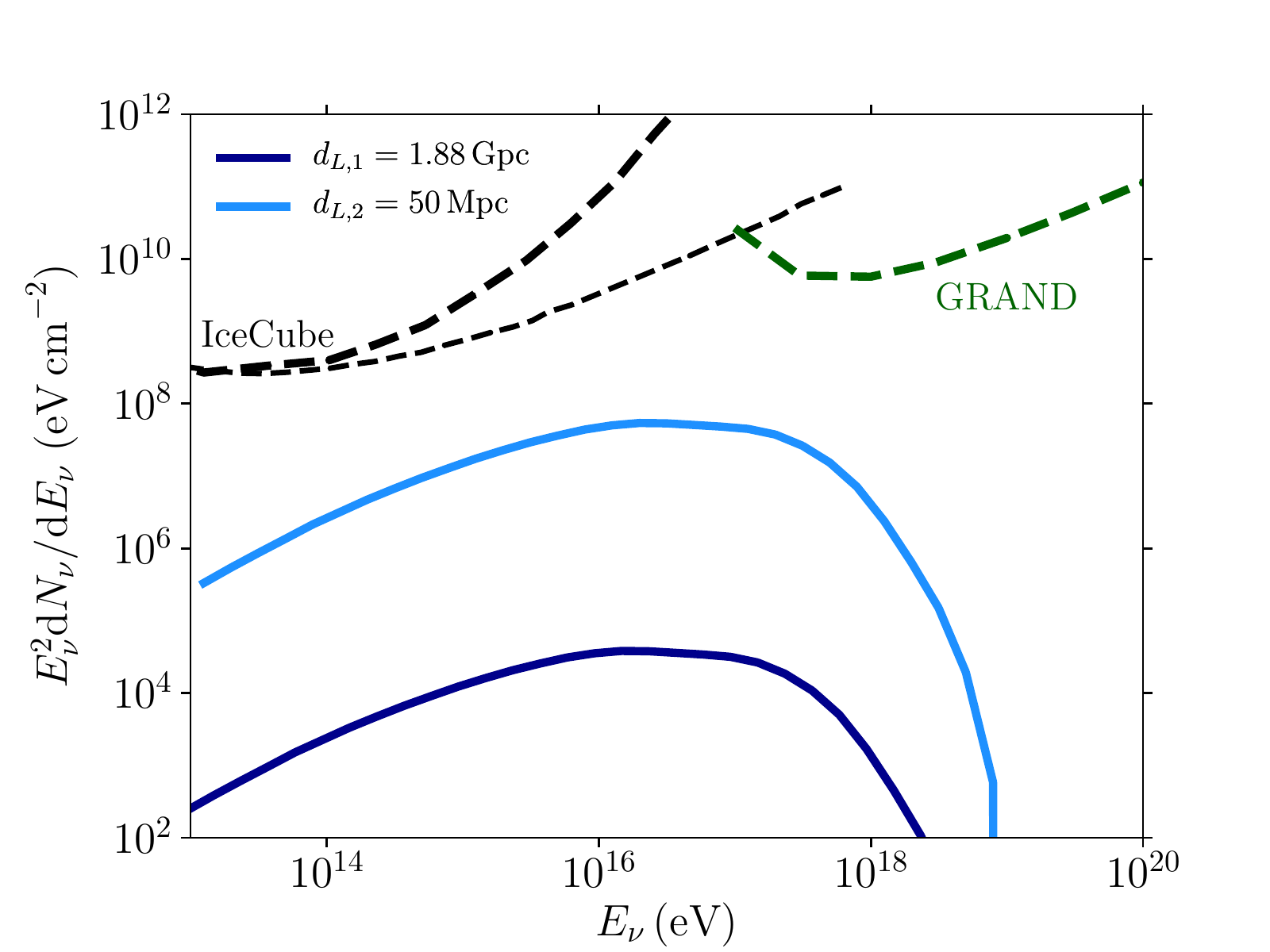}
\caption{Neutrino spectra for three flavors for one source with same characteristics as in Fig.~\ref{Fig:CR_Fe_1e46_007_110_10}. We show the total spectra (in ${\rm eV\,cm}^{-2}$) for a high state SED with $L_{\rm pk}=10^{47.5}\,{\rm erg\,s}^{-1}$ (left) and a medium state SED with $L_{\rm pk}=10^{46}\,{\rm erg\,s}^{-1}$ (right). We consider two different distances $d_{L,1} =1.88\,{\rm Gpc}$ (dark blue) and $d_{L,2} =50\,{\rm Mpc}$ (light blue). The IceCube and projected GRAND sensitivities \citep{Fang17ICRC} are also shown (dashed black and green lines). For the IceCube sensitivities, we show two cases depending on the declination: $0\degree<\delta<30\degree$ (most favorable case, thin line) and $30\degree<\delta<60\degree$ (Swift J1644+57 case, thick line) \citep{Aartsen14}.
}\label{Fig:Nu_Fe_1e46_007_110_10}
\end{figure*}

We show in Fig.~\ref{Fig:CR_Fe_1e46_007_110_10} an example of outgoing cosmic-ray spectrum for a pure iron injection from a single TDE in its high state SED characterized by $L_{\rm pk}=10^{47.5}\,{\rm erg\,s}^{-1}$ and $\hat{a}=0.07$, and in its medium state SED characterized by $L_{\rm pk}=10^{46}\,{\rm erg\,s}^{-1}$ and $\hat{a}=0.07$. As is shown below, these two states are associated in our model with a black hole of mass $M_{\rm bh} = 7 \times 10^{6} M_\odot$. We consider an injection spectral index of $\alpha=1.8$ and an acceleration efficiency $\eta_{\rm acc}=0.2$. Here we do not account for the extragalactic propagation of cosmic rays, and the spectrum is normalized by considering the luminosity distance of Swift J1644+57: $d_{ L,1} \simeq 1.88 \, {\rm Gpc}$ ($z \simeq 0.354$). Two associated neutrino spectra are shown in Fig.~\ref{Fig:Nu_Fe_1e46_007_110_10}. One spectrum is normalized by considering the luminosity distance of Swift J1644+57, $d_{ L,1} \simeq 1.88 \, {\rm Gpc}$, and the other by considering a luminosity distance $d_{L,2} = 50 \, {\rm Mpc}$. The IceCube sensitivity is characterized by a minimum fluence $ {\cal S}_{\rm IC} = 5 \times 10^{-4}$ TeV cm$^{-2}$ over the energy range 10\,TeV$-$10\,PeV, which corresponds to a detection limit $ s_{\rm IC} \sim 10^{-11}$ TeV cm$^{-2}$ s$^{-1}$ for a one-year data collection \citep{Aartsen15}. We give the IceCube sensitivity from the effective area presented in \cite{Aartsen14} for the optimal declination range $0\degree<\delta<30\degree$ (thin lines), and for the declination range  $30\degree<\delta<60\degree$ (thick lines) associated with the Swift event J1644+57.

The peak luminosity and width of the photon SED have a strong effect on the cosmic-ray and neutrino spectra as they influence strongly photohadronic and synchrotron losses, which are the two dominant energy loss processes in our framework. For cosmic rays, energy  losses due to photohadronic interactions are mainly dominant at low energies, while synchrotron losses dominate at high energies. If the radiation energy density is sufficiently low, the escape time of cosmic rays can be the limiting time at low energies.

Regarding the cosmic-ray spectrum, we see in Fig.~\ref{Fig:CR_Fe_1e46_007_110_10} that for a medium state SED with $L_{\rm pk}=10^{46}\,{\rm erg\,s}^{-1}$, iron strongly interacts and produces many secondary particles with a large number of nucleons below $E_{\rm cut}/56 \sim 10^{19}\,{\rm eV}$. Despite these high interaction rates, nuclei can still survive and escape from the region with energies up to $10^{20}\,{\rm eV}$. For a high state SED with $L_{\rm pk}=10^{47.5}\,{\rm erg\,s}^{-1}$, the iron strongly interacts, as do the secondary cosmic rays produced through iron interactions. No nuclei can survive and escape the region; only protons escape, with a maximum energy around $E_{p,{\rm max}} \sim 10^{18}\,{\rm eV}$. The high-energy cutoff for each element with $Z>1$ results from the competition between the energy loss processes (see Appendix~\ref{Appendix::Emax}) or from the maximum injection energy; in Fig.~\ref{Fig:CR_Fe_1e46_007_110_10}, acceleration is the limiting process for $L_{\rm pk} = 10^{46}\,{\rm erg\,s}^{-1}$, and photonuclear interactions for $L_{\rm pk} = 10^{47.5}\,{\rm erg\,s}^{-1}$.

Figure~\ref{Fig:Nu_Fe_1e46_007_110_10} shows that a nearby medium state TDE at distance $50\,{\rm Mpc}$ with peak luminosity $L_{\rm pk}=10^{46}\,{\rm erg\,s}^{-1}$ would not be detectable, even with future neutrino detectors such as GRAND \citep{Fang17ICRC}. On the other hand, at early times and in their high states, TDEs would lead to massive production of high-energy neutrinos, and should be marginally detectable with IceCube and with GRAND at the high-energy end for a nearby distance of $50\,${Mpc}. We note that the rate of TDEs at distances smaller than $50\,${Mpc} is $4 \times 10^{-6}$ yr$^{-1}$ for a comoving event rate density $ \sim 0.03 \, {\rm Gpc}^{-3}\, {\rm yr}^{-1}$, which is extremely low. TDEs in their high states at distances $>50\,{\rm Mpc}$ would not be detectable with IceCube or GRAND because of the flux decrease and the low high-energy cutoff of the neutrino spectrum. We note that our chosen parameter set is consistent with the non-detection of neutrinos from Swift J1644+57 (as was already highlighted in \citealp{Guepin17}) and allows for baryonic loading at the source $\xi_{\rm CR}$ up to $\sim$ a few 100 to remain consistent with this non-detection. 

The presence of a plateau in the neutrino spectrum is due to the contributions of muon and electron neutrinos. The high-energy cutoff is due to pions and muons experiencing energy losses (mainly synchrotron losses) before they decay. We account for synchrotron and inverse Compton losses, but do not account for the kaon contribution.
We note that electron neutrinos have a lower energy cutoff than muon neutrinos. Electron neutrinos are produced through muon decay, and muons are produced through pion decay; therefore, the energy of electron neutrinos is influenced by pion losses and muon losses before they decay. Muon neutrinos, in turn, are produced through pion decay or muon decay. Hence the energy of those produced through pion decay is only influenced by pion energy losses before the decay, which explains the higher energy cutoff for muon neutrinos.

\section{Modeling the population of TDEs contributing to UHECR and neutrino fluxes}
\label{section:population}

A derivation of the comoving density rate of TDEs can be found in \cite{Sun15}. These authors define the comoving density rate as $\dot{n}(z,L) = \dot{n}_0  \Lambda(L) f(z)$, where $\dot{n}_0$ is the total local event rate density, $f(z)$ the TDE redshift distribution, and $\Lambda(L)$ the TDE luminosity function. The luminosity function is given by a power-law,
\begin{equation}
\Lambda_{\rm TDE}(L_\gamma) \propto \left(\frac{L_{\gamma,{\rm pk}}}{L_{\rm m,pk}}\right)^{-\alpha_L}  \, ,
\end{equation}
with $L_{\rm m,pk} = 10^{48}\,{\rm erg \,s}^{-1}$ and $\int_{L_{\rm min}}^{L_{\rm max}} {\rm d}L_\gamma \, \Lambda_{\rm TDE}(L_\gamma) = 1$, with $L_{\rm min} = 10^{45}\,{\rm erg \,s}^{-1}$ and $L_{\rm max} = 10^{49}\,{\rm erg \,s}^{-1}$, and $\alpha_L=2$.

However, the  \cite{Sun15} model is not well adapted to our framework as their comoving rate density accounts for the entire TDE population and not the subpopulation powering jets. Moreover, the redshift evolution of the luminosity function is neglected, due to the small size of their observational sample.

Thus, in the following we present a prediction for the comoving event rate density of TDEs powering jets by combining the TDE rate per galaxy $\dot{N}_{\rm TDE}$ and the black hole comoving number density per luminosity  ${\rm d}n_{\rm bh}(z,L)/{\rm d}L$   (i.e., the number of black holes per comoving volume and per bin of jet luminosity).

\subsection{TDE rate per galaxy}

The TDE rate per galaxy $\dot{N}_{\rm TDE}$ depends on the galaxies considered. Following \cite{Wang04}, we consider a lower bound in the case of core galaxies of
\begin{equation}
\dot{N}_{\rm TDE} \approx 10^{-5}\,{\rm yr}^{-1} \, ,
\end{equation}
and an upper bound in the case of power-law galaxies of
\begin{equation}\label{Eq:TDE_rate2}
\dot{N}_{\rm TDE} \approx 7.1 \times 10^{-4} \,{\rm yr}^{-1} \, \left(\frac{\sigma}{70\,{\rm km\,s}^{-1}} \right)^{7/2} M_{{\rm bh},6}^{-1} \, ,
\end{equation}
where $M_{{\rm bh},6} = M_{\rm bh}/10^6 M_\odot$ and $\sigma$ is the stellar velocity dispersion of the bulge. From \cite{Kormendy13}, the relation between the black hole mass and the bulge velocity dispersion is
\begin{equation}\label{Eq:M_sigma}
\log_{10} M_{{\rm bh},9} = -0.51 + 4.4 \log_{10}\left( \frac{\sigma}{200\,{\rm km\,s}^{-1}} \right) \, .
\end{equation}
Using Eqs.~\ref{Eq:TDE_rate2} and \ref{Eq:M_sigma}, we obtain the TDE rate per galaxy in the case of power-law galaxies, which depends only on the black hole mass:

\begin{equation}
\dot{N}_{\rm TDE} \approx 3 \times 10^{-4} \,{\rm yr}^{-1}\, M_{{\rm bh},6}^{-0.2} \, .
\end{equation}

\subsection{Identifying the black hole masses leading to observable TDEs}\label{Section:BHmass}
First, we identify the population of black holes that can lead to observable TDEs.  
TDEs can occur for stellar objects of mass $M_\star$ at distances $d_\star \leq r_{\rm t} = R_\star (M_{\rm bh}/M_\star)^{1/3}$ \citep{Hills75}. Following \cite{Krolik12}, we can estimate the tidal disruption radius $r_{\rm t} \simeq 10  R_{\rm s}\, M_{\star,\odot}^{2/3-\xi} M_{{\rm bh},6}^{-2/3} \left[(k_\star/f_\star)/0.02\right]^{1/6} $, where $R_{\rm s}= 2GM_{\rm bh}/c^2$ is the Schwarzschild radius, $M_{\star,\odot}$ is the mass of the star in solar units, $M_{{\rm bh},6}=M_{{\rm bh}}/10^6 M_\odot$, $k_\star$ is related to the star's radial density profile, and $f_\star$ is its binding energy in units of $GM_\star^2/R_\star$. This radius is obtained for a main sequence star with a typical mass--radius relation $R_\star \approx R_\odot M_{\star,\odot}^{1-\xi}$ with $\xi \simeq 0.2$ for $0.1 M_\odot < M_\star \leq M_\odot$ or $\xi \simeq 0.4$ for $M_\odot < M_\star < 10 M_\odot$ \citep{Kippenhahn94}. Moreover, we  consider here and in what follows fully radiative stars, thus $k_\star/f_\star = 0.02$ \citep{Phinney89}. For white dwarfs, typically $R_\star\sim M_\star^{-1/3}$ with $0.5 M_\odot < M_\star \leq 0.7\,M_\odot$. Their tidal disruption radii are smaller due to the smaller dimensions of these objects; an approximate formula gives $r_{\rm t} \simeq 7.4 \times 10^9 \, (M_{{\rm bh},3.3}/\rho_{\star,7})^{1/3}$ \citep{Luminet89}, where $\rho_\star$ is the white dwarf core density.

With this tidal disruption radius, we can estimate the maximum black hole mass enabling the production of flares. The first-order requirement for flares to be produced reads $r_{\rm t} \gtrsim R_{\rm s}$. For a Schwarzschild black hole, this leads to $M_{\rm bh} \lesssim 4 \times 10^7 M_\odot \, M_{\star,\odot}^{1-3\xi/2}  \left[(k_\star/f_\star)/0.02 \right]^{1/4}$, which ranges from $M_{\rm bh} \lesssim 10^7 M_\odot $ to $M_{\rm bh} \lesssim 10^8 M_\odot $ for $0.1 M_\odot < M_\star \leq 10 M_\odot$. However, jetted TDEs are likely to be powered by black holes with moderate to high spin; a general-relativistic treatment accounting for the black hole spin increases the maximum mass of black holes that are able to disrupt a solar-like star: $M_{\rm bh} \sim 7 \times 10^8 M_\odot $ \citep{Kesden12}.

\subsection{Relation between black hole mass and jet luminosity}
The black hole mass function ${\rm d}n_{\rm bh}(z,M_{\rm bh})/{\rm d}M_{\rm bh}$  (i.e., the number of black holes per comoving volume and per mass bin) is obtained with the semi-analytic galaxy formation model review in Sect. \ref{Section::MBH}, and we derive  ${\rm d}n_{\rm bh}(z,L)/{\rm d}L$ by relating the black hole mass and the jet luminosity. Following \cite{Krolik12}, we consider a TDE which forms a thick accretion disk, powering a jet through the Blandford--Znajek mechanism. We estimate the maximum accretion rate by considering that about $1/3$ of the stellar mass is accreted after one orbital period $P_{\rm orb}$ \citep{Lodato09a}. From \cite{Krolik12},
\begin{equation}
P_{\rm orb}(a_{\rm min}) \approx  5 \times 10^{5}\,{\rm s}\, M_{\star,\odot}^{(1-3\xi)/2} M_{{\rm bh},6}^{1/2} \left(\frac{k_\star/f_\star}{0.02} \right)^{1/2} \beta^3 \, ,
\end{equation}
where $a_{\rm min}$ is the minimum semi-major axis. The parameter $\xi$ comes from the main sequence mass--radius relation $R_\star \approx R_\odot M_{\star,\odot}^{(1-\xi)}$ and $\beta \lesssim 1$ is the penetration factor. We obtain the following accretion rate:
\begin{equation}
\dot{M} \approx  20 M_\odot \,{\rm yr}^{-1} M_{\star,\odot}^{(1+3\xi)/2} M_{{\rm bh},6}^{-1/2} \left(\frac{k_\star/f_\star}{0.02} \right)^{-1/2} \beta^{-3}\, .
\end{equation}
The luminosity of a jet powered by a black hole depends on the regime of accretion. In the super-Eddington regime, i.e., for $M_{\rm bh} \lesssim M_{\rm bh,jet}$, where $M_{\rm bh,jet} = 4 \times 10^8 M_\odot \, (\dot{m}/\dot{m}_0)^{2/3} M_{\star,\odot}^{(1+\xi)/3} \left[ (k_\star/f_\star)/0.02 \right]^{-1/2}  \beta^{-3} $, the jet luminosity is given by \cite{Krolik12},
\begin{equation}\label{Eq:L_M}
\begin{split}
L_{\rm jet} \approx & \; 10^{43} \,{\rm erg\, s}^{-1} \, \frac{f(a)}{\beta_h \alpha_{\rm s}} M_{{\rm bh},6}^{-1/2}  \times 8 \times 10^3 \, q (\dot{m}/\dot{m}_0) \\
& \times  M_{\star,\odot}^{(1+3\xi)/2} \left( \dfrac{k_\star/f_\star}{0.02}\right)^{-1/2}  \beta^{-3}\, ,
\end{split}
\end{equation}
 where $\alpha_{\rm s}$ is the ratio of inflow speed to orbital speed of the disk, and $\beta_h$ the ratio of the midplane total  pressure near the  ISCO to the magnetic pressure in the black hole's stretched horizon, such that $\alpha_{\rm s}\beta_h \sim 0.1-1$; the function $f(a)\approx a^2$ encodes the dependence of the jet luminosity on the dimensionless spin of the black hole~\citep{2015MNRAS.453..157P}, $a$, which ranges from $a=0$ (for a Schwarzschild black hole) to $a=1$ (for a maximally spinning black hole); $\dot{m} = \dot{M}c^2/L_{\rm Edd}$ is the normalized accretion rate in the outer disk (with $L_{\rm Edd}$ the Eddington luminosity); $\dot{m}_0$ is the peak normalized accretion rate; and $q$ is the fraction of $\dot{m}$ arriving at the black hole, thus accounting for possible outflows. We do not consider the sub-Eddington regime as it involves black holes with higher masses, which should not be able to tidally disrupt main sequence stars. We recall that in the following we assume default values $q=1$, $a=1$, $\alpha_{\rm s}\beta_h=1$, $k_\star/f_\star=0.02$ for the parameters appearing in the jet luminosity. In particular, the choice to set the spin $a=1$ is justified by iron-$K\alpha$ measurements of AGN spins~\citep{2013mams.book.....B,2013CQGra..30x4004R}, on which our galaxy formation model is calibrated~\citep{sesana2014}. Clearly, if all black holes had low spins $a\ll1$ our jetted TDE rates would be significantly decreased, but such a choice seems hard to reconcile with iron-$K\alpha$ observations (see discussion in \citealt{sesana2014}).

The total energy release per TDE is given by
\begin{equation}
{\mathcal E}_{\rm jet}\approx L_{\rm jet}P_{\rm orb}\simeq 4 \times{10}^{52}~{\rm erg}~M_{\star,\odot}\,\frac{f(a)}{\beta_h \alpha_{\rm s}} q (\dot{m}/\dot{m}_0) \, ,
\end{equation}
which should be less than $\dot{M}c^2$. 
We note that a jet luminosity $L_{\rm jet}=\eta_{\rm jet}\dot{M}c^2$ and $\eta_{\rm jet}\sim1$ are achieved if the disk is magnetically arrested, but the efficiency factor may be smaller. Also, the gravitational binding energy is much lower, so we need to rely on an energy extraction, for example via the Blandford--Znajek mechanism to have powerful jets. 

The observable non-thermal luminosity (which, as mentioned before, is identified with the bolometric luminosity of our SED model) is related to the jet luminosity by accounting for the efficiency of energy conversion from Poynting to photon luminosity $\eta_{\rm jet}$ and for the beaming factor $\mathcal{B} = \min(4\pi/\Delta\Omega,2\Gamma^2)$, where $\Delta\Omega$ is the solid angle occupied by the jet \citep{Krolik12}. For a two-sided jet with a jet opening angle $\theta_{\rm jet}$, we have $\mathcal{B} = \min(1/(1-\cos\theta_{\rm jet}),2\Gamma^2)$. Therefore, $L_{\rm bol} = L_{\rm jet, obs} = 2 \,\eta_{{\rm jet},-2} \mathcal{B} L_{\rm jet}$ for $\theta_{\rm jet} \sim 5\degree$, $\Gamma = 10$, and $\eta_{\rm jet} = 0.01$.

Considering the theoretical local rate density $\dot{n}_{{\rm tde},0} = 150 \,{\rm Gpc}^{-3}\,{\rm yr}^{-1}$ estimated from the semi-analytic galaxy formation model of \cite{Barausse2012} (see section~\ref{Section::MBH}), the luminosity density is estimated to be
\begin{align}
Q_{\rm TDEjet}&\approx \eta_{\rm jet}\,\mathcal{B}\,{\mathcal E}_{\rm jet}\,\dot{n}_{{\rm tde},0} \,,\nonumber\\
 &\simeq {10}^{46}~{\rm erg}~{\rm Mpc}^{-3}~{\rm yr}^{-1} \, \eta_{{\rm jet},-2}  M_{\star,\odot}\,\frac{f(a)}{\beta_h \alpha_{\rm s}} \nonumber\\& \times  q \left(\frac{\dot{m}}{\dot{m}_0}\right)\frac{\dot{n}_{{\rm tde},0}}{ 150 \,{\rm Gpc}^{-3}\,{\rm yr}^{-1}}\, .
\end{align}
We note that $\sim 10^{43}\,{\rm erg}~{\rm Mpc}^{-3}~{\rm yr}^{-1}$ above $10^{19.5}\,{\rm eV}$ is required to account for the observed flux of UHECR for a hard injection spectral index \citep[e.g.][]{Katz09}. This estimate can be modified for a heavy composition due to the contribution of secondary cosmic rays below $10^{20}\,{\rm eV}$. However, the uncertainties remain high on the local event rate density given the small number of jetted TDEs observed.

\subsection{Redshift evolution of the black hole mass function}
\label{Section::MBH}
To model the cosmological evolution of massive black holes in their galactic hosts, we utilize the semi-analytic galaxy formation 
model of \cite{Barausse2012} \citep[with incremental improvements described in][]{sesana2014,Antonini2015,Antonini_Barausse2015,Bormio_PTA1,Bormio_PTA2}, adopting
the default calibration of \citet{2017MNRAS.468.4782B}. The model describes the cosmological evolution of baryonic structures on
top of Dark Matter merger trees produced with the extended Press--Schechter formalism, modified to more closely reproduce the results of
N-body simulations within the $\Lambda$CDM model~\citep{Press1974,Parkinson2008}. Among the baryonic structures that are evolved along the branches of the merger trees, and which
merge at the nodes of the tree, are a diffuse, chemically primordial intergalactic medium, either shock-heated to the Dark Matter halo's virial temperature or 
streaming into the halo in cold filaments (the  former  is more common at low redshift and high halo masses, and the latter in small systems at high redshifts;~\citealt{Dekel2006,Cattaneo2006,Dekel2009});
a cold interstellar medium (with either disk- or bulge-like geometry), which forms from the cooling of the intergalactic medium or from the above-mentioned cold accretion flows, and which can give rise to star formation in a quiescent fashion or in bursts~\citep{sesana2014}; pc-scale nuclear star clusters, forming from the migration of globular clusters to the galactic center or by in situ star formation~\citep{Antonini2015,Antonini_Barausse2015}; a central massive black hole, feeding from a reservoir of cold gas, brought to the galactic center, for example by major mergers and disk bar instabilities. Our semi-analytic model also accounts for feedback processes on the growth or structures (namely from supernovae and from the jets and outflows produced by AGNs), and for the sub-pc evolution of massive black holes,  for example the evolution of black hole spins~\citep{Barausse2012,sesana2014},  migration of binaries due to gas interactions, stellar hardening and triple massive black hole interactions~\citep{Bormio_PTA1,Bormio_PTA2},
gravitational-wave emission~\citep{Klein2016}. 

For the purposes of this paper, the crucial input provided by our model is the evolution of the
TDE luminosity function. We determine the TDE comoving rate density $\dot{n}_{\rm TDE}(z,L)$ by combining the TDE rate per galaxy $\dot{N}_{\rm TDE}$ and the black hole comoving density $n_{\rm bh}(z,M_{\rm bh})$, and using the black hole mass and jet luminosity relation (Eq.~\ref{Eq:L_M}).

For the properties of the jet, we distinguish between the high state, characterized by a high jet efficiency, and the medium state, characterized by a lower jet efficiency. We set these parameters in order to be consistent with the observations of Swift J1644+57, which should be associated with a black hole of mass $M_{\rm bh} \gtrsim 7 \times 10^6 \,M_\odot$ \citep{Seifina17} and reaches a bolometric luminosity  $L_{\rm bol} \gtrsim 10^{48}\,{\rm erg\,s}^{-1}$ in the high state. Therefore, we have $\theta_{\rm jet} = 5\degree$ and $\eta_{\rm jet} = 0.35$ in the high state and $\theta_{\rm jet} = 5\degree$ and $\eta_{\rm jet} = 0.01$ in the medium state. 
For instance, a black hole of mass $M_{\rm bh} \gtrsim 7 \times 10^6 \,M_\odot$ is associated with $L_{\rm bol} \simeq 2 \times 10^{48}\,{\rm erg\,s}^{-1}$ in the high state and $L_{\rm bol} \simeq 7 \times 10^{46}\,{\rm erg\,s}^{-1}$ in the medium state. The other cases that we consider are shown in Table~\ref{Tab::Mbh_Ljet}. As explained in section~\ref{Section:BHmass}, for high masses $M_{\rm bh} > 10^8 \,M_\odot$, only highly spinning black holes could lead to observable flares. For completeness, we also account for this case in our study.  The black hole mass functions at different redshifts and detailed comparisons of the predictions of our model to observational determinations are given in Appendix~\ref{App:BHz}.

\begin{table}[t]
\begin{center}
\caption{Observed jet luminosities as a function of black hole mass $M_{\rm bh}\,(M_\odot)$ in the medium state $L_{ \rm bol, med}\, ( {\rm erg\,s}^{-1})$ for $\theta_{\rm jet} = 5\degree$ and $\eta_{\rm jet} = 0.01$, and in the high state $L_{\rm bol, high}\,({\rm erg\,s}^{-1})$ for $\theta_{\rm jet} = 5\degree$ and $\eta_{\rm jet} = 0.35$.}\label{Tab::Mbh_Ljet}
\begin{tabular}{cccc}
\toprule  $M_{\rm bh}$  & $L_{ \rm bol, med}\,[L_{\rm pk}]$ & $L_{\rm bol, high}\,[L_{\rm pk}]$ \\
$(M_\odot)$  & $( {\rm erg\,s}^{-1})$ & $({\rm erg\,s}^{-1})$ \\
\midrule
\midrule  $7 \times 10^8$  &  $6.8 \times 10^{45} [10^{45}]$ & $2.1 \times 10^{47}[10^{46.5}]$  \\
\midrule  $7 \times 10^7$  &  $2.1 \times 10^{46} [10^{45.5}]$ & $6.8 \times 10^{47}[10^{47}]$  \\
\midrule  $7 \times 10^6$  &  $6.8 \times 10^{46} [10^{46}]$ & $2.1 \times 10^{48}[10^{47.5}]$  \\
\midrule  $7 \times 10^5$  &  $2.1 \times 10^{47} [10^{46.5}]$ & $6.8 \times 10^{48}[10^{48}]$  \\
\bottomrule
\end{tabular}
\end{center}
\end{table}

It is interesting to notice that the luminosity function of jetted TDEs is dominated by high luminosities (hence low black hole masses) in our model, unlike the distribution of \cite{Sun15}. This stems from the flat black hole mass functions at low masses (Figs.~\ref{MF}) combined with the $L_{\rm jet}\propto M_{\rm bh}^{-1/2}$ relation (Eq.~\ref{Eq:L_M}). It implies, quite naturally, which of the observed very bright objects such as Swift J1644+57 are the dominant ones in the population. These objects thus set the maximum bolometric luminosity $L_{\rm max}$ in the luminosity function, which we introduce as a cutoff in our population model. This also implies that the diffuse flux of UHECRs will be dominantly produced by the most luminous objects in their medium state.

\begin{figure}[t]
\centering
\includegraphics[width=0.49\textwidth]{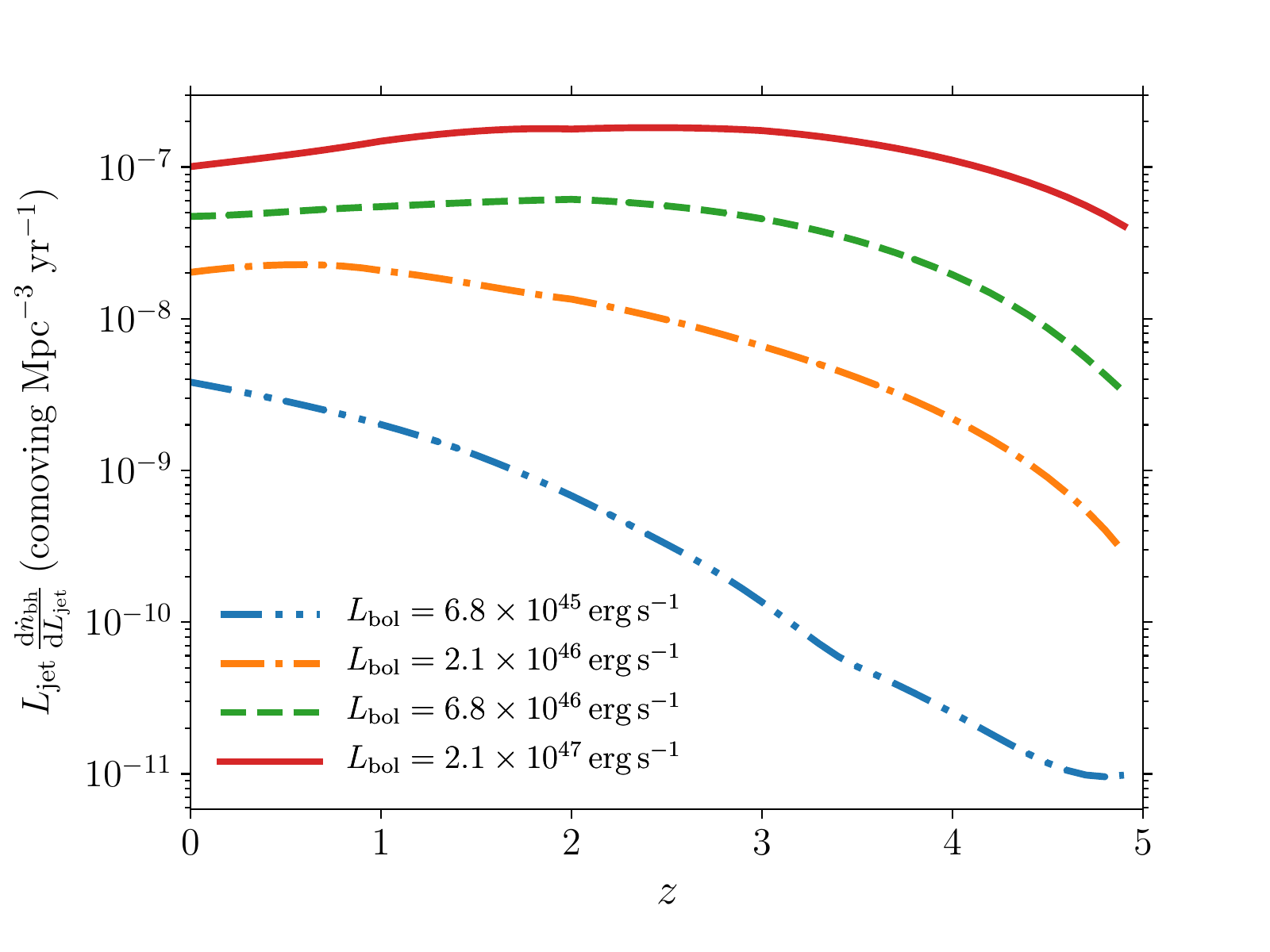}
\caption{Comoving TDE luminosity density in their medium state as a function of redshift, as derived in our model, for $\dot{N}_{\rm TDE} = 10^{-5}\,{\rm yr}^{-1}$. The different luminosities correspond to different black hole masses.}
\label{Fig:BHz}
\end{figure}

Figure~\ref{Fig:BHz} shows that the corresponding TDE comoving rate density remains almost constant up to redshift $\sim 3$ for luminosities $\ge 10^{45.5}\,$erg\,s\,$^{-1}$, which dominate in the production of cosmic-ray and neutrino fluxes in our framework.

\section{Diffuse UHECR and neutrino fluxes from a TDE population}\label{section:diffuse}

In the following we calculate the diffuse cosmic-ray and neutrino fluxes, and the composition of cosmic rays by considering a population of jetted TDEs. All primed quantities are in the jet comoving frame, all quantities with superscript $c$ are in the source comoving frame, and all other quantities are in the observer frame. The fluxes depend on the spectra produced by each source, on the comoving rate of TDEs (detailed in the previous subsections), and on the cosmic-ray propagation to the Earth. During the extragalactic propagation, cosmic rays may interact with the cosmic microwave background (CMB) and the extragalactic background light (EBL) through photonuclear interactions. Because of these processes, they may lose energy and create secondary particles in the case of nuclei. In our work we consider EBL models from \cite{Kneiske04} and \cite{Stecker06}.

The diffuse cosmic-ray flux is given by
\begin{equation}
\resizebox{0.9\hsize}{!}{$
\begin{split}
\Phi_{\rm CR}(E_{\rm CR}) = \frac{c}{4 \pi H_0} \int\limits_{z_{\rm min}}^{z_{\rm max}} \int\limits_{L_{\rm min}}^{L_{\rm max}} &{\rm d}z \,{\rm d}L\, \frac{{ f_{\rm s}\,{ \xi_{\rm CR}}\,}\dot{N}_{\rm TDE} \,{\rm d}n_{\rm bh}(z,L)/{\rm d}L}{\sqrt{\Omega_{\rm M}(1+z)^3+\Omega_\Lambda}}   \\ & \times F^c_{{\rm CR,s,p}}(E^c_{\rm CR},z,L) t^c_{\rm dur} \,,
\end{split}
$}
\end{equation}
where $\Omega_{\rm M}=0.3$ and  $\Omega_{\rm L}=0.7$ are our fiducial cosmological parameters, $H_0 = 70 \,{\rm km \,s}^{-1}\,{\rm Mpc}^{-1}$ is the Hubble constant, $F^c_{{\rm CR,s,p}}(E^c_{\rm CR},z,L)$ is the spectrum obtained after the propagation of cosmic rays from a source at redshift $z$ (per bin of comoving enegy $E^c_{\rm CR}$ and per unit of comoving time $t^c$), and $t^c_{\rm dur}$ is the duration of the emission in the source comoving frame. The TDE rate $\dot{N}_{\rm TDE}$ is the rate of TDEs per galaxy and ${\rm d}n_{\rm bh}(z,L)/{\rm d}L$ is the comoving black hole density per (jet) luminosity bin;  $f_{\rm s}$ is the fraction of the jetted TDE population, calculated in Sect.~\ref{section:population},  which contributes to the production of UHECRs.  

Similarly, the diffuse neutrino flux reads
\begin{equation}
\resizebox{0.9\hsize}{!}{$
\begin{split}
\Phi_\nu(E_\nu) = \frac{c}{4 \pi H_0} \int\limits_{z_{\rm min}}^{z_{\rm max}} \int\limits_{L_{\rm min}}^{L_{\rm max}} &{\rm d}z \,{\rm d}L\, \frac{ f_{\rm s}\,{ \xi_{\rm CR}}\dot{N}_{\rm TDE} \,{\rm d}n_{\rm bh}(z,L)/{\rm d}L}{\sqrt{\Omega_{\rm M}(1+z)^3+\Omega_\Lambda}}  \\ & \times F^c_{\nu,{\rm s}}(E^c_\nu,L) t^c_{\rm dur} \,,
\end{split}
$}
\end{equation}
where $F^c_{\nu,{\rm s}}(E^c_\nu,L)$ is the neutrino flux, per comoving energy and per comoving time, for a source with luminosity $L$.

Due to the flat evolution of the TDE comoving density rates up to $z\sim 3$, we can safely use the jetted TDE luminosity distribution at $z=0$ in the above equations and separate the integrals in $L$ and $z$. We checked in particular that our results were similar when using the distribution function at redshifts $z\lesssim 3$. The impact of the redshift evolution on the cosmic-ray spectrum and on the neutrino flux level is also limited, and close to a uniform evolution as described in \cite{Kotera10}.

\subsection{Final spectrum and composition of cosmic rays}\label{sec:diffuse_UHECR}

\begin{figure}[t]
\centering
\includegraphics[width=0.49\textwidth]{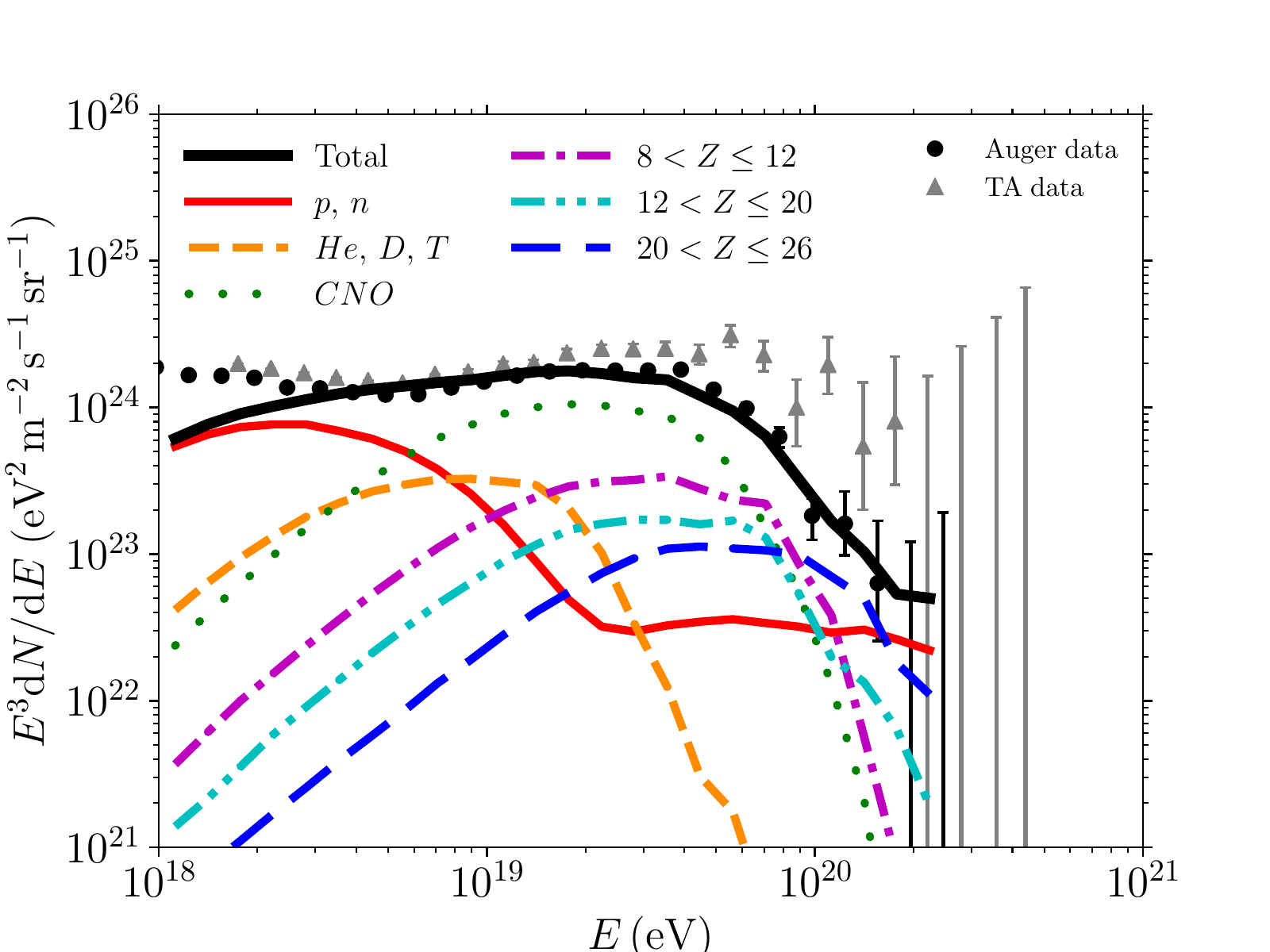}
\caption{ Diffuse cosmic-ray spectrum from a population of jetted TDEs (calculated in their medium states) obtained for an injection of 70\% Si and 30\% Fe, with spectral index $\alpha=1.5$, $\xi_{\rm CR}f_{\rm s} = 2.6 \times 10^{-3}$ and source evolution derived in this work, with maximum bolometric luminosity in the population $L_{\rm max}=6.8\times10^{46}\,$erg\,s$^{-1}$. We show the total spectrum (black) and its composition. We superimpose data from the Auger experiment (black dots, \citealp{Auger15}), and from the Telescope Array experiment (gray dots, \citealp{TA_Fukushima15}) for which only statistical uncertainties are shown.}\label{Fig:Tot_spec}
\end{figure}

We show in Fig.~\ref{Fig:Tot_spec} the cosmic-ray spectrum obtained for an injection of 70\% Si and 30\% Fe, a spectral index $\alpha=1.5$, an acceleration efficiency $\eta_{\rm acc} = 0.1$, a TDE source evolution, and a fraction $\xi_{\rm CR}f_{\rm s} = 2.6 \times 10^{-3}$ of the local event rate density $\dot{n}_{{\rm tde},0} = 1.5 \times 10^2 \,{\rm Gpc}^{-3}\,{\rm yr}^{-1}$. This rate is computed from the TDE rate per galaxy obtained in the case of core galaxies. The population fraction corresponds approximately to the rate density $\sim 0.4\,{\rm Gpc}^{-3}\,{\rm yr}^{-1}$. This heavy composition could be injected for example by the core of disrupted stars. We recall that we consider a production of UHECRs dominated by medium states.

Superimposed are the data from the Auger experiment \citep{Auger15} and from the Telescope Array experiment \citep{TA_Fukushima15} shown with their statistical uncertainties. We note that the systematic uncertainty on the energy scale is $14\%$ for the Auger experiment.

\begin{figure}[t]
\centering
\includegraphics[width=0.49\textwidth]{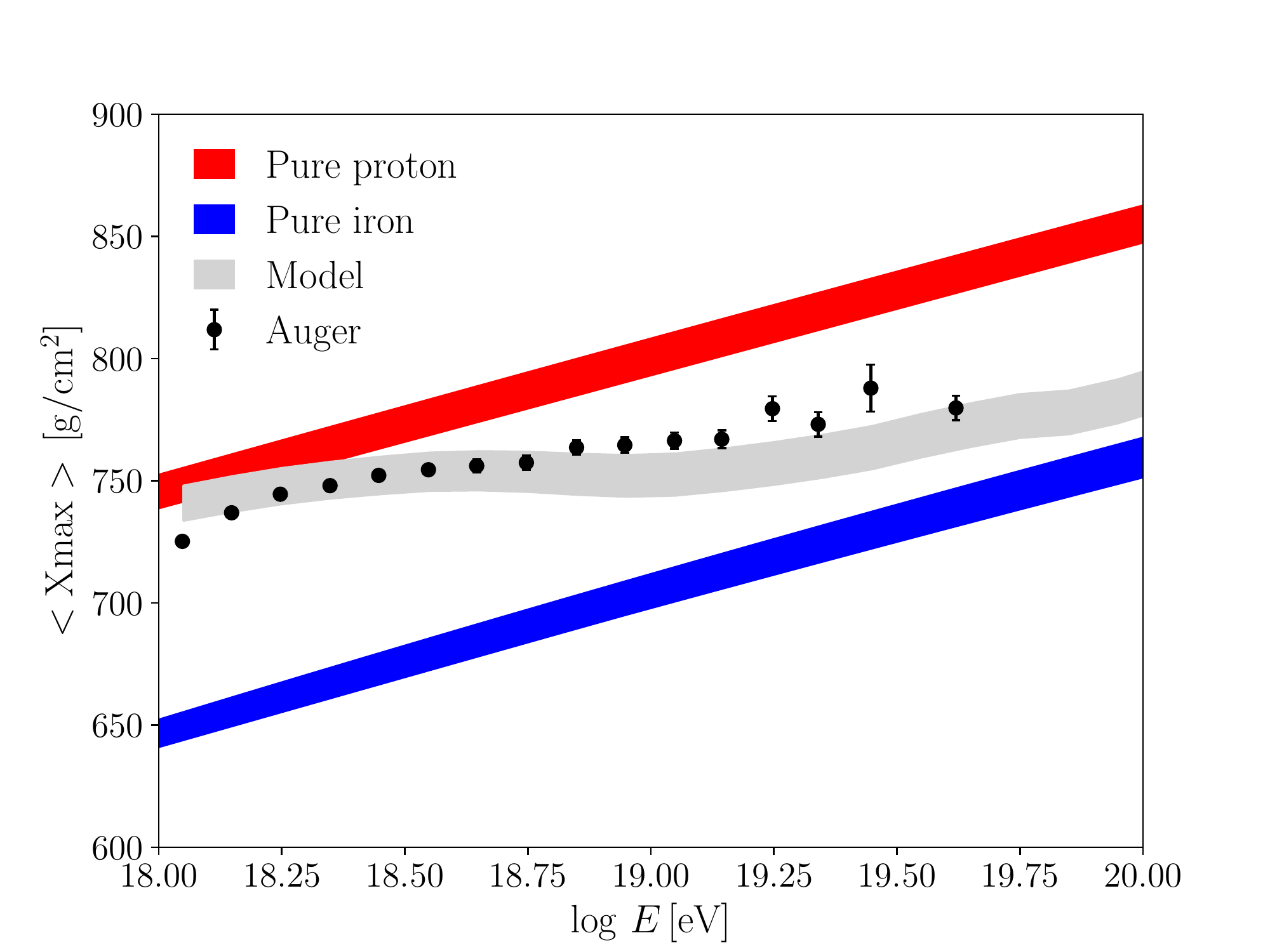}
\includegraphics[width=0.49\textwidth]{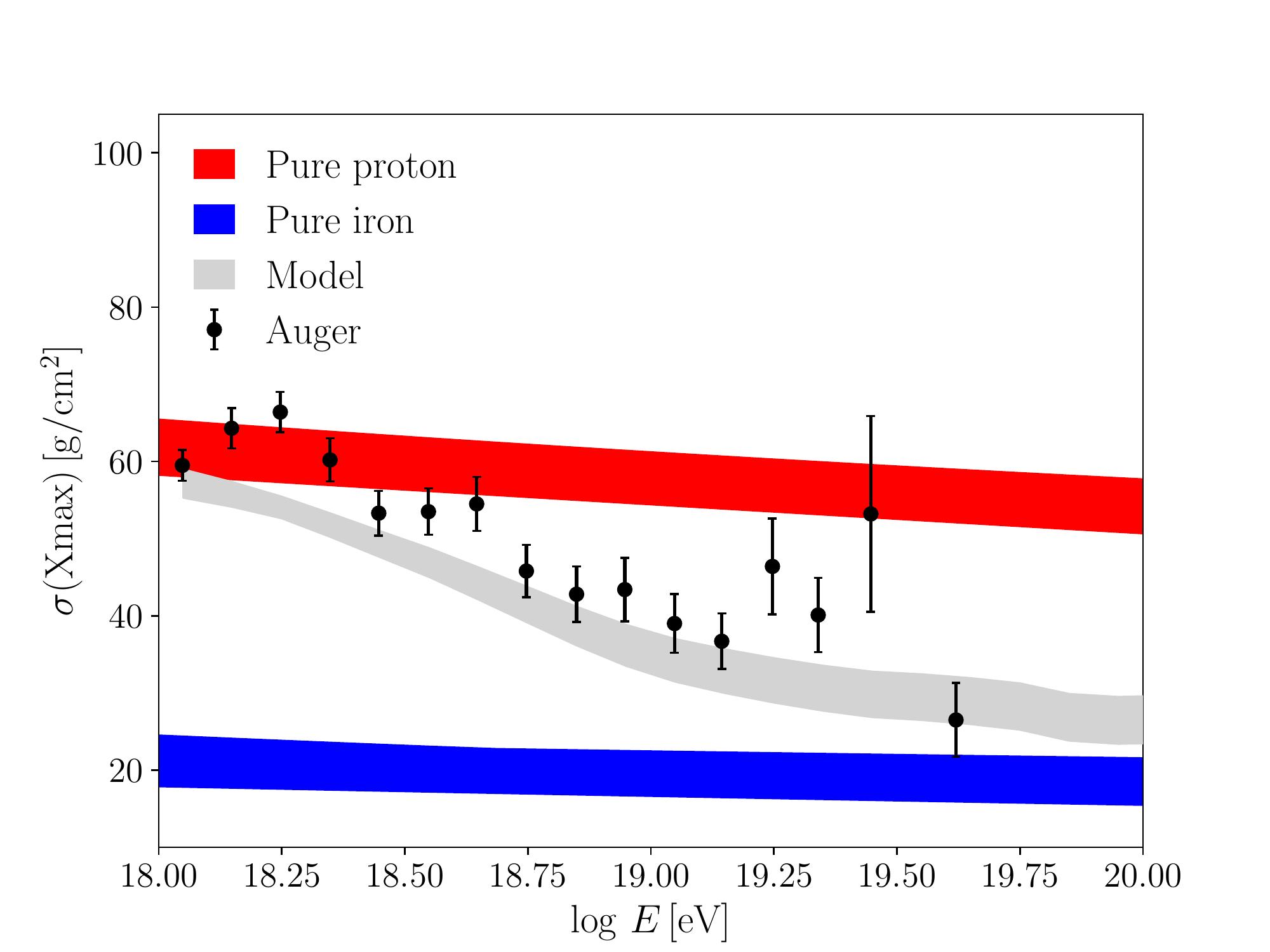}
\caption{Mean and standard deviation of $X_{\rm max}$ for the spectrum shown in Fig.~\ref{Fig:Tot_spec} (pale gray band). We also show Auger measurements \citep{Auger_Xmax15} with uncertainties (black dots) and simulation results for pure proton injection (red band) and pure iron injection (blue band). The bands are obtained by accounting for hadronic model uncertainties ({\sc Epos-LHC}, {\sc Sibyll 2.1} or {\sc QGSJet II-04}). }\label{Fig:Tot_compo}
\end{figure}

We also show in Fig.~\ref{Fig:Tot_compo}  the corresponding mean and standard deviation for the depth of the maximum of the air showers, $\left\langle X_{\rm max} \right\rangle$ and $\sigma(X_{\rm max})$. It is represented by a gray band, due to uncertainties related to UHECR-air interaction models {\sc Epos-LHC} \citep{Werner06,Pierog13}, {\sc Sibyll 2.1} \citep{Ahn09}, or {\sc QGSJet II-04} \citep{Ostapchenko06b,Ostapchenko13}. Superimposed are the data from the Auger experiment for the composition of UHECRs, $\left\langle X_{\rm max} \right\rangle$ and $\sigma(X_{\rm max})$ \citep{Auger_Xmax15}, which are in good agreement with our model. Only statistical uncertainties are displayed; systematic uncertainties are at most $\pm 10\,{\rm g\,cm}^{-2}$ for $\left\langle X_{\rm max} \right\rangle$ and $\pm 2\,{\rm g\,cm}^{-2}$ for $\sigma(X_{\rm max})$.
Our results are compatible with a light composition at $10^{18.5}\,{\rm eV}$, shifting toward a heavier composition for increasing energy.

With increased cutoff bolometric luminosities $L_{\rm max}$, harder injection spectra are needed in order to compensate for the abundant production of nucleons at low energies, which softens the overall spectrum (typically $\alpha=1$ is required for $L_{\rm max}=10^{47}\,$erg\,s$^{-1}$). We present the case of the injection of a dominant fraction of heavy elements; the injection of more intermediate elements, such as the CNO group, is possible at the cost of increasing the acceleration efficiency $\eta_{\rm acc}$, and hardening the injection spectrum further in order to achieve the highest energies.

Our model allows for a fit of both the UHECR spectrum and composition of the Auger observations, as long as the dominant sources supply luminosities $\lesssim 10^{46.5}\,$erg\,s$^{-1}$, which is a value that is consistent with the observed Swift event. We note that if the high states were dominant for the UHECR production, we would not be able to fit the data; because of the strong photodisintegration of heavy elements in the very dense radiation background, we would obtain a large production of nucleons below $10^{19}\,{\rm eV}$ and no survival of heavy elements at the highest energies. However, because of its limited duration ($t_{\rm dur} \sim 10^5\,{\rm s}$ is chosen as an upper bound), the high state is unlikely to be dominant. In a more refined model we should account for the evolution of the luminosity of the jet, which should decrease during the event duration.

The disrupted stellar object provides material (protons and heavier nuclei) that can be injected and accelerated in a jet. As already emphasized, the composition of this material is poorly constrained; it could be similar to the composition of the stellar object or modified during the disruption process. It is interesting that white dwarfs could be commonly disrupted by intermediate mass black holes. These stars could be a source of copious amounts of CNO nuclei, which seem to be observed in the composition of UHECRs measured by Auger, as noted in \cite{AlvesBatista17}. For completeness, we  tested in this study various injection fractions, and we  present one case that allows us to fit  the Auger data well. We note that a deviation of 5\% in the composition does not largely affect the fit to the data within the error bars,  given the uncertainties on the other jetted TDE parameters.

Markers of the occurrence of jets associated with TDEs were detected only very recently. Most TDEs should not power jets and only a small fraction of jetted TDEs should point toward the observer, depending on the jet opening angle. Therefore, the properties of these objects are still subject to large uncertainties. From an observational perspective, the jetted TDEs detected recently are very luminous events with a peak isotropic luminosity $L_{\rm pk} \sim 10^{47}-10^{48}\,{\rm erg\,s}^{-1}$, and a local event rate density is of $\dot{n}_{{\rm tde},0} \sim 0.03 \,{\rm  Gpc}^{-3}\, {\rm yr}^{-1}$ \citep[e.g.,][]{farrar2014}. On the other hand, normal TDEs are less luminous and are characterized by a higher local event rate density $\dot{n}_{{\rm tde},0} = 10^2\,{\rm Gpc}^{-3}\,{\rm yr}^{-1}$ \citep[e.g.,][]{donley02}. However, the characteristics of this new population, mainly their luminosity distribution and comoving event rate density,  are difficult to infer due to the scarcity of observations.  From our population model, the maximum local event rate density that we can expect reaches $\dot{n}_{{\rm tde},0} \sim 2 \times 10^2 \,{\rm Gpc}^{-3}\,{\rm yr}^{-1}$ for core galaxies and $\dot{n}_{{\rm tde},0} \sim 3 \times 10^3 \,{\rm Gpc}^{-3}\,{\rm yr}^{-1}$ for power-law galaxies. 

The fraction needed to fit the UHECR spectrum of the Auger observations, $\xi_{\rm CR}f_{\rm s} = 2.6 \times 10^{-3}$, can account for example for low UHECR injection rates $\xi_{\rm CR}$, and/or for population constraints, such as the fraction of TDE jets pointing toward the observer. Assuming the low rate inferred from the observations of $\dot{n}_{{\rm tde},0} \sim 0.03 \,{\rm  Gpc}^{-3}\, {\rm yr}^{-1}$ for the jetted events pointed toward the observer, a baryon loading of $\xi_{\rm CR}\sim 10$ is required. This value is consistent with the non-detection of neutrinos from Swift J1644+57, which implies an upper limit to the baryon density of a few 100 \citep{Senno16b,Guepin17}. 

At the lowest energies, the high state could contribute marginally to the diffuse cosmic-ray flux, as shown in figure~\ref{Fig:Nu_Fe_1e46_007_110_10}. The strong photodisintegration in the high state leads to a strong production of nucleons below $10^{19}\,{\rm eV}$, which would add a new component to the spectrum and make the composition lighter. This would lead to a better fit of the composition in Fig.~\ref{Fig:Tot_compo}.

\subsection{Diffuse neutrino flux}

\begin{figure}[!tp]
\centering
\includegraphics[width=0.49\textwidth]{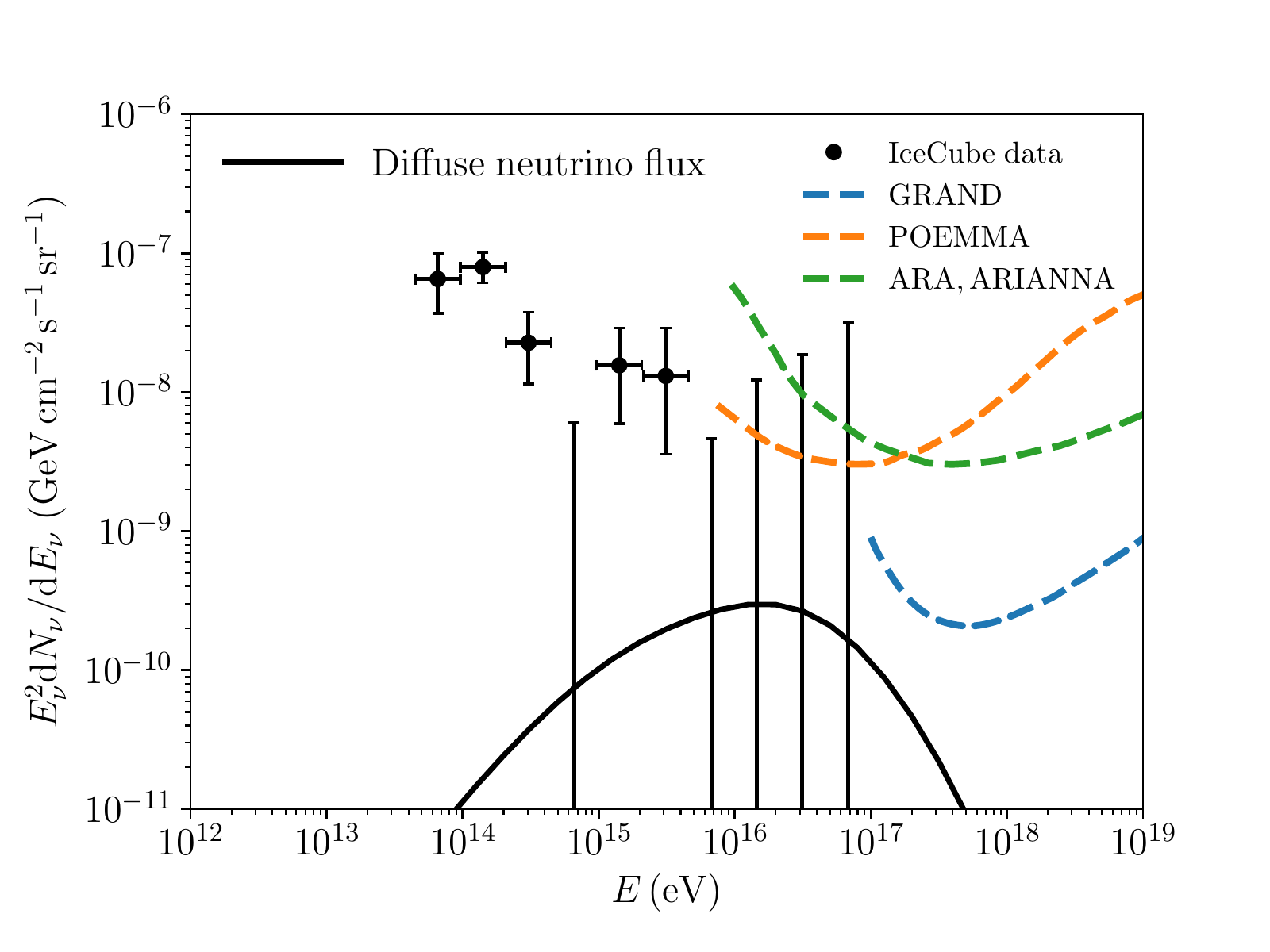}
\caption{ Diffuse neutrino flux for three flavors from a population of jetted TDEs with the same properties as in Fig.~\ref{Fig:Tot_spec} (neutrino production calculated in their high states). We also show the diffuse neutrino flux measured by the IceCube experiment \citep{IC17}, and the projected limits for GRAND \citep{Fang17ICRC}, ARA/ARIANNA \citep{ARA15,ARIANNA15}, and POEMMA \citep{Neronov17}. }\label{Fig:Tot_spec_nu}
\end{figure}

The TDE event density obtained by fitting the Auger data with our UHECR spectrum allows us to calculate the diffuse neutrino flux from a population of TDEs, by considering the fraction $\xi_{\rm CR}f_{\rm s}$ calculated above. As shown in Fig.~\ref{Fig:Tot_spec_nu}, the diffuse neutrino flux from jetted TDEs contributes marginally to the total diffuse neutrino flux observed by IceCube \citep{IC17}. As it peaks at high energies, around $10^{16}\,{\rm eV}$, it could be a good target for future generation detectors. However, we note that this flux is too low to be detectable with  ARA/ARIANNA, POEMMA, and GRAND at even higher energies.  Its high-energy cutoff reduces the flux at higher energies, and therefore it lies below the GRAND sensitivity limit.

\section{Discussion and conclusion}

We assessed in this study the production of UHECRs and neutrinos by a population of TDEs. In our model, the disruption of a stellar object launches a relativistic jet, where internal shocks can accelerate a part of the disrupted material, namely  light and heavy nuclei. This scenario is connected to recent observations and analytic studies, favoring a jetted model for some very luminous events. In such a case, material from the disrupted object can be injected and accelerated inside the jet, and can experience interactions before escaping and propagating in some cases toward the Earth. However, other scenarios could be contemplated; for instance, a substantial fraction of the accreted material could be ejected as a wind where particles could be linearly accelerated\footnote{\cite{Zhang17} show that UHECR acceleration is difficult in the wind model.}. 

The bulk Lorentz factor $\Gamma$ and the opening angle of the jet $\theta_{\rm jet}$ are two additional important quantities impacting our results. The bulk Lorentz factor of the jet impacts the observed jet isotropic luminosity and the energy of detected cosmic rays and neutrinos. The dynamical time also scales as $\Gamma$ and the photon energy density as $\Gamma^{-4}$, thus an increase of a factor of a few in $\Gamma$ could lead to a drastic cut in the photodisintegration rates. Here we use the fiducial value $\Gamma = 10$, but for larger values we  expect that the survival of nuclei would be favored, leading to lower nucleon production at lower energies, and thus to a larger parameter space allowing for a good fit for the diffuse UHECR spectrum. Our choice is conservative in this sense. On the other hand, the neutrino production would be consequently reduced. The opening angle of the jet is also not well constrained; therefore, we adopt a small value $\theta_{\rm jet} \sim 5\degree$ for the high and medium states \citep[e.g.,][]{Bloom2011,Burrows2011}. 
Like the bulk Lorentz factor, this parameter is also involved in the model that we use to link the black hole mass to the isotropic luminosity of the jet. The jet can be seen only if it is pointing toward the observer. However, we note that the effective opening angle for cosmic rays might be higher than the usual opening angle as cosmic rays can experience small deflections inside the jet; thus, misaligned jetted TDEs characterized by a higher rate than aligned events might also contribute to the diffuse cosmic-ray flux.

While finalizing this paper, we became aware of the independent work of \cite{Biehl17} on a similar topic. 
These authors show that the acceleration of nuclei in jets created by the tidal disruption of white dwarfs can lead to a simultaneous fit of the UHECR data and the measured IceCube neutrino flux in the PeV range. One major difference with our study is that we include a detailed jetted TDE population study by modeling the luminosity function and rate evolution in redshift. Our conclusions are also different from theirs, in so far as we cannot fit the observed diffuse IC neutrino flux with our TDE population model. 
This negative result is consistent with several arguments already highlighted in previous works by \cite{Dai16} and \cite{Senno16b}. In particular, the absence of observed neutrino multiplets in the IceCube data gives a lower limit of $\gtrsim 100-1000\,{\rm Gpc}^{-3}\,{\rm yr}^{-1}$, which is significantly higher than the rate of jetted TDEs pointing toward us inferred from observations of $\sim 0.03 \,{\rm  Gpc}^{-3}\, {\rm yr}^{-1}$, and  higher than  rates with dimmer luminosities  also constrained by X-ray observations. In addition, large baryon loadings with $\gtrsim 1000$ are ruled out as such values would imply that Swift J1466+57 would have been observed in neutrinos. Also, a large baryon loading factor implies a total TDE energy of $\gtrsim10^{54}$~erg, which violates the energetics argument. 
We emphasize that considering high an medium states has a significant impact on the fact that we cannot reproduce both the observed UHECR and HE neutrino diffuse fluxes with our model. Most of the UHECRs contributing to the diffuse flux are produced during the medium state, whereas most of the HE neutrinos are produced during the high state. Considering only the high state would require to increase the baryon loading or the TDE event rate in order to reproduce the observed UHECR spectrum, and would thus increase the associated HE neutrino flux.
 
Our model is able to reproduce with a reasonable accuracy and for a reasonable range of parameters the observations from the Auger experiments, and TDE powering jets  therefore appear to be  good candidates for the production of UHECRs. Our results are consistent with other TDE studies that also obtain good fits to UHECR data: \cite{Zhang17} stress that oxygen-neon-magnesium white-dwarf TDE models could provide good fits, but do not account for photodisintegration in the vicinity of the source because they used a steeper luminosity function. Our model can account for these interactions, and allows us to explore the parameter space for the radiation field, the injection and the composition. This is important for our flatter luminosity function, which predicts that the luminosity density is dominated by the highest-power TDEs, i.e., the effective luminosity is $L\sim 2 \times 10^{48}~{\rm erg}~{\rm s}^{-1}$ in the high state.

The associated transient HE neutrinos could be detected for single nearby sources (at distances of a few tens of Mpc) with IceCube and upcoming instruments at higher energies such as GRAND or POEMMA. The diffuse flux would be within reach of IceCube in the next decade. Its detection would be more challenging for future generation instruments aiming at the detection of ultra-high-energy neutrinos, due to a high-energy cutoff below $10^{17}\,$eV. 

Among the other transient UHECR nuclei models that have been suggested to explain the UHECR data (e.g., fast rotating pulsars, \citealp{FKO12,Fang14,Kotera15} or GRBs, \citealp{WangGRB08,Murase08,Globus15_GRB,Globus15}), the jetted TDE model has the interesting property of presenting two different states (medium and high)  leading to optimal production of both UHECRs and neutrinos. In addition, the jetted TDE scenario appears mildly constrained by photon observations. Within our model, we demonstrated that the observed Swift J1466+57 can be seen as a typical source that would dominate the production of UHECRs and neutrinos. Even under this constraint, the wide range of variation allowed for several free parameters (for example the Lorentz factor of the outflow, as discussed earlier) enables us to correctly fit the cosmic-ray data. A specific signature of this scenario is thus difficult to infer. A direct multi-messenger signal with TDE photons associated with the emission of neutrinos from a single source appears to be the way to validate this scenario.

\section*{Acknowledgements}
We thank Tanguy Pierog for his continuous help on hadronic interactions and EPOS. We are indebted to Marta Volonteri for enlightening conversations on the physics of massive black holes. We also thank Joe Silk, Rafael Alves Batista, and Nicholas Senno for very fruitful discussions. This work is supported by the APACHE grant (ANR-16-CE31-0001) of the French Agence Nationale de la Recherche. CG is supported by a fellowship from the CFM Foundation for Research and by the Labex ILP (reference ANR-10-LABX-63, ANR-11-IDEX-0004-02). EB is supported by the European Union's Horizon 2020 research and innovation
program under the Marie Sklodowska-Curie grant agreement No. 690904. For our simulations, we have made use of the Horizon Cluster, hosted by the
Institut d'Astrophysique de Paris. We thank Stephane Rouberol for running  this cluster smoothly for us. The work of KM is supported by the Alfred P. Sloan Foundation and NSF grant No. PHY-1620777.

\bibliographystyle{aa}
\bibliography{TDEbib}

\appendix

\section{Cosmic-ray maximum energy}\label{Appendix::Emax}

We derive analytic estimates of the effective maximum energies in the comoving frame of the jet as a function of the energy loss processes at play. They depend on the characteristics of the event considered, and its radiation region, for example the bolometric luminosity $L_{\rm bol}$, the comoving mean magnetic field $B'$, the time variability $t_{\rm var}$, or the bulk Lorentz factor $\Gamma$.

\subsection{Maximum injection energy}
As explained in Sect.~\ref{Sec::model}, the maximum injection energy $E'_{Z, \rm max}$ of a nucleus of charge $Z$ is determined by the competition between the acceleration timescale $t'_{\rm acc} = \eta_{\rm acc}^{-1} E'/ c\,Z\,e\,B'$ and the energy loss timescales $t'_{\rm loss}= \min(t'_{\rm dyn},t'_{\rm syn}, t'_{\rm IC}, t'_{\rm BH}, t'_{p\gamma},...)$. An upper bound of the maximum energy of accelerated particles is thus given by the competition between the acceleration timescale ($t'_{\rm acc}$) and the dynamical timescale ($t'_{\rm dyn}$):
\begin{eqnarray}
E'_{ Z, \rm up, \rm dyn} &\sim & c \,Z\, e \, B' \, (1+z)^{-1} \Gamma \,t_{\rm var} \,\eta_{\rm acc} \, , \\
&\simeq & 6.3 \times 10^{17} \,{\rm eV} \, Z_1 \,B'_{2.8} \, \Gamma_{10} \, t_{{\rm var},2} \,\eta_{{\rm acc},-1} \, ,\nonumber\\
&\simeq & 1.6 \times 10^{19} \,{\rm eV} \, Z_{26} \,B'_{2.8} \, \Gamma_{10} \, t_{{\rm var},2} \,\eta_{{\rm acc},-1} \, .\nonumber
\end{eqnarray}

The competition between the acceleration timescale ($t'_{\rm acc}$) and the synchrotron timescale ($t'_{\rm syn}$) gives

\begin{eqnarray}
E'_{ Z, \rm up, \rm syn} &\sim & \left[ \frac{6\pi(m_{\rm u} c^2)^2 e }{(m_e/m_{\rm u})^2 \sigma_{\rm T}} \right]^{\frac{1}{2}} A^2 Z^{-3/2} B'^{-1/2} \eta_{\rm acc}^{1/2} \, ,\\
&\simeq & 2.4 \times 10^{18} \,{\rm eV} \, A^2_1\,Z^{-3/2}_1 \,B'^{-1/2}_{2.8} \, \,\eta^{1/2}_{{\rm acc},-1} \, ,\nonumber\\
&\simeq & 5.7 \times 10^{19} \,{\rm eV} \, A^2_{56}\,Z^{-3/2}_{26} \,B'^{-1/2}_{2.8} \, \,\eta^{1/2}_{{\rm acc},-1} \, .\nonumber
\end{eqnarray}

Here, the mean magnetic field ${B'} = \sqrt{8\pi \int {\rm d}\epsilon' \, \epsilon'n'_{\epsilon'}} \simeq 10^{2.85}\,{\rm G}$ is obtained for a log-parabola SED with peak luminosity $L_{\rm pk} = 10^{46}\,{\rm erg\,s}^{-1}$ and a width $\hat{a} = 0.07$.

The upper bound given by the competition between the acceleration timescale ($t'_{\rm acc}$) and the photohadronic timescale ($t'_{\rm p\gamma}$) is computed numerically. For the parameters considered above, we obtain for protons $E'_{ p, \rm up, \rm p\gamma} \simeq 2.5 \times 10^{18} \,{\rm eV}$. The comparison between the different energy loss timescales allows us to determine the limiting energy loss process: for the previous example the dynamical timescale is the limiting timescale.

\subsection{Competition between energy loss processes in the radiation region}
The competition between the energy loss processes in the radiation region influences the outgoing cosmic-ray spectrum, and in particular the high-energy cutoffs. By considering the competition between synchrotron losses ($t'_{\rm syn}$) and escape ($t'_{\rm esc}$) for protons,
\begin{equation}
 \frac{6\pi(m_p c^2)^2}{(m_e/m_p)^2 \sigma_{\rm T} c E'_p {B'}^2}= \Gamma t_{\rm var} \, ,
\end{equation}
with $E'_p$ the proton energy in the comoving frame, $\sigma_{\rm T}$ the Thomson cross section, $m_p$ the proton mass, $m_e$ the electron mass, and $c$ the speed of light, we can derive the corresponding high-energy cutoff:
\begin{eqnarray}\label{Eq:Epmax}
 E'_{p,{ \rm max} } &=& \frac{6\pi(m_p c^2)^2}{(m_e/m_p)^2 \sigma_{\rm T} c {B'}^2 \Gamma t_{\rm var}} \, ,\\
 & \simeq &  9.0\times 10^{18} \,{\rm eV} \, {B'}^{-2}_{2.8} \Gamma^{-1}_1 t_{{\rm var}, 2}^{-1}\, . \nonumber
\end{eqnarray}

The competition between synchrotron losses ($t'_{\rm syn}$) and pion production ($t'_{p\gamma}$) for a transient event characterized by a hard spectrum gives
\begin{equation}
 \frac{6\pi(m_p c^2)^2}{(m_e/m_p)^2 \sigma_{\rm T} c E'_p {B'}^2}= \frac{4\pi \Gamma^5 c^2 t_{\rm var}^2 \epsilon_{\rm pk}} {\left\langle \sigma_{p\gamma} \kappa_{p\gamma} \right\rangle L_{\rm pk}} \, ,
\end{equation}
with $\sigma_{p\gamma}$ and $\kappa_{p\gamma}$ the photopion cross section and inelasticity, $\epsilon_{\rm pk}$ the peak energy, and $L_{\rm pk}$ the peak luminosity. For $\epsilon_{\rm pk} = 70\,{\rm keV}$, we obtain a high-energy cutoff:
\begin{eqnarray}
 E'_{ p,{\rm max} }&=& \frac{3 (m_p c^2)^2 \left\langle \sigma_{p\gamma} \kappa_{p\gamma} \right\rangle L_{\rm pk}}{2 (m_e/m_p)^2 \sigma_{\rm T} c^3 {B'}^2 \Gamma^5 t_{\rm var}^2 \epsilon_{\rm pk}} \, ,\\
 & \simeq &  5.0\times 10^{21} \,{\rm eV} \, {B'}^{-2}_{2.8} \, \Gamma^{-5}_1  L_{{\rm pk},46} \, t_{{\rm var}, 2}^{-2}\, . \nonumber
\end{eqnarray}
Our numerical estimates are evaluated for typical parameters of jetted TDEs (e.g., the characteristics of Swift J1644+57).

For nuclei, the high-energy cutoffs depend on the mass and atomic numbers. As $\gamma_{N} = E'_{N}/ A m_{\rm u} c^2$, where $m_{\rm u}$ is the atomic mass unit,  for the competition between synchrotron losses ($t'_{\rm syn}$) and escape ($t'_{\rm esc}$) we obtain
 \begin{eqnarray}
 E'_{N, { \rm max}} &=& \frac{6\pi(m_{\rm u} c^2)^2 (A/Z)^4}{(m_e/m_{\rm u})^2 \sigma_{\rm T} c {B'}^2 \Gamma t_{\rm var}} \, ,\\
 & \simeq &  1.4\times 10^{20} \,{\rm eV} \, {B'}^{-2}_{2.8} \Gamma^{-1}_1 t_{{\rm var}, 2}^{-1}\,A_{56}^4\,Z_{26}^{-4} , \nonumber
\end{eqnarray}
where $A_{56}=A/56$ and $Z_{26}=Z/26$ for iron nuclei.

\section{Derivation of diffuse neutrino and cosmic-ray fluxes}\label{Appendix:Diffuse_Fux}

To calculate the neutrino diffuse flux, we integrate the neutrino flux of a single source over the TDE population. Primed quantities are in the jet comoving frame, quantities with superscript $c$ are in the source comoving frame, and other quantities are in the observer frame. We account for the total number of neutrinos produced by one single source, which depends on the neutrino energy in the source comoving frame $E^c_\nu = (1+z) E_\nu$ and the bolometric luminosity of the source: $N_{\nu,{\rm s}}(E^c_{\nu},L)$.

Moreover, for a redshift $z$, we can observe a population of sources characterized by a comoving event rate density $\dot{n}(z,L)$ during an observation time $T_{\rm obs} = (1+z) T^c_{\rm obs}$ (where $T_{\rm obs}$ is the time in the observer frame and $T^c_{\rm obs}$ is the time in the source comoving frame), in a comoving volume,
\begin{equation}
\frac{{\rm d}V(z)}{{\rm d}z} = \frac{c}{H_0}\frac{4\pi D_{\rm c}^2}{\sqrt{\Omega_{\rm M}(1+z)^3+\Omega_\Lambda}}\,,
\end{equation}
where $\Omega_{\rm M}=0.3$ and  $\Omega_{\rm L}=0.7$ are our fiducial cosmological parameters, and $H_0 = 70 \,{\rm km \,s}^{-1}\,{\rm Mpc}^{-1}$ is the Hubble constant. The comoving event rate density ${\rm d}\dot{n}(z,L)/{\rm d}L = \dot{N}_{\rm TDE} \,{\rm d}n_{\rm bh}(z,L)/{\rm d}L$ depends on the TDE rate per galaxy $\dot{N}_{\rm TDE}$ and the black hole comoving density per luminosity ${\rm d}n_{\rm bh}(z,L)/{\rm d}L$. Therefore, the diffuse neutrino flux is given by
\begin{equation}
\resizebox{0.85\hsize}{!}{$
\begin{split}
\frac{{\rm d}N_\nu}{{\rm d}E_\nu}(E_\nu) = &\frac{1}{4\pi} \int\limits_{z_{\rm min}}^{z_{\rm max}} \int\limits_{L_{\rm min}}^{L_{\rm max}} \left[{\rm d}z \,{\rm d}L \,{ f_{\rm s}\, \xi_{\rm CR}} \frac{{\rm d}\dot{n}(z,L)}{{\rm d}L} \right.\\ & \times \left. \frac{1}{1+z}\, \frac{{\rm d}V(z)}{{\rm d}z} \, \frac{1}{4\pi D_{\rm c}^2}\frac{{\rm d}N_{\nu,{\rm s}}(E^c_{\nu},L)}{{\rm d}E^c_{\nu}} \frac{{\rm d} E^c_{\nu}}{{\rm d}E_{\nu}}\right] \, .
\end{split}
$}
\end{equation}
As $E^c_{\nu} = (1+z) E_{\nu}$, we obtain a  diffuse neutrino flux 
\begin{equation}
\resizebox{0.88\hsize}{!}{$
\begin{split}
\Phi_\nu(E_\nu) = \frac{c}{4 \pi H_0} \int\limits_{z_{\rm min}}^{z_{\rm max}} \int\limits_{L_{\rm min}}^{L_{\rm max}} &{\rm d}z \,{\rm d}L\, \frac{{ f_{\rm s}\, \xi_{\rm CR}}{\rm d}\dot{n}(z,L)/{\rm d}L}{\sqrt{\Omega_{\rm M}(1+z)^3+\Omega_\Lambda}} \\ & \times F^c_{\nu,{\rm s}}(E^c_\nu,L) t^c_{\rm dur} \,,
\end{split}
$}
\end{equation}
where $\Phi_\nu(E_\nu) = {\rm d}^2 N_{\nu}(E_{\nu})/{\rm d}E_{\nu} {\rm d}t $ is the diffuse neutrino flux per unit time (observer frame), $F^c_{\nu,{\rm s}}(E^c_\nu,L) = {\rm d}^2N_{\nu,{\rm s}}(E^c_{\nu},L)/{\rm d}E^c_{\nu} {\rm d}t^c$ is the number of neutrinos emitted by one single source per bin of comoving energy and per unit of comoving time, and $t^c_{\rm dur}$ is the duration of the emission in the source comoving frame.

The cosmic-ray diffuse flux is calculated in a similar manner, but in this case we also need to account for the large-scale propagation of cosmic rays
\begin{equation}
\resizebox{0.88\hsize}{!}{$
\begin{split}
\Phi_{\rm CR}(E_{\rm CR}) = \frac{c}{4 \pi H_0} \int\limits_{z_{\rm min}}^{z_{\rm max}} \int\limits_{L_{\rm min}}^{L_{\rm max}} &{\rm d}z \,{\rm d}L\, \frac{{ f_{\rm s}\, \xi_{\rm CR}}{\rm d}\dot{n}(z,L)/{\rm d}L}{\sqrt{\Omega_{\rm M}(1+z)^3+\Omega_\Lambda}} \\ & \times F^c_{{\rm CR,s,p}}(E^c_{\rm CR},z,L) t^c_{\rm dur} \,,
\end{split}
$}
\end{equation}
where $F^c_{{\rm CR,s,p}}(E^c_{\rm CR},z,L)$ is the spectrum obtained after the propagation of cosmic rays from a source at redshift $z$.

\section{Evolution of the black hole mass function in redshift}\label{App:BHz}

\begin{figure*}[!tp]
\centering
\includegraphics[width=0.9\textwidth]{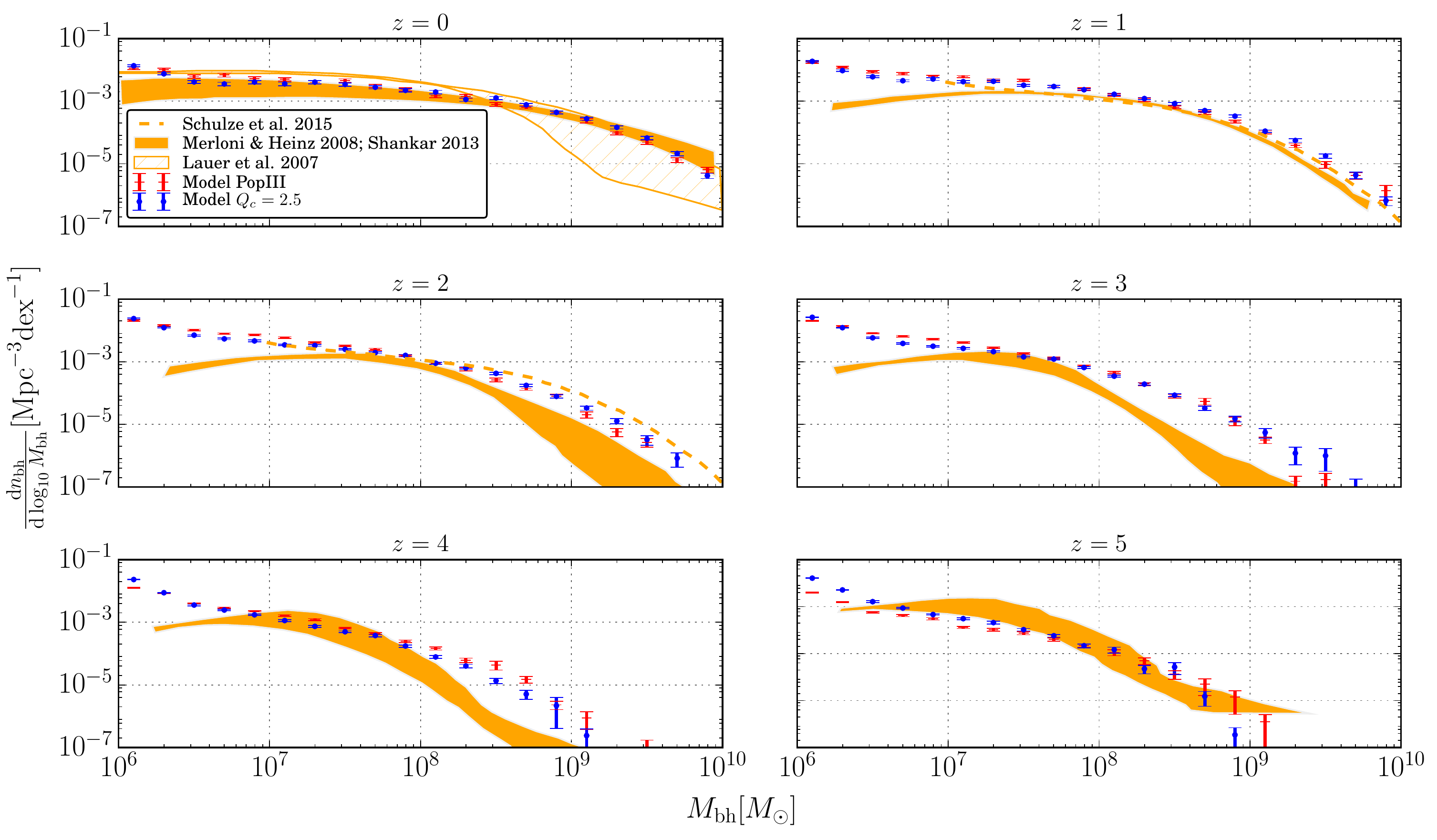}
\caption{Predictions of our model in two different scenarios,  light-seed  (``Pop III'') and  heavy-seed  (``$Q_c=2.5$''), for the mass function of black holes as a function of redshift. The model's error bars are Poissonian. For comparison, also shown are the observational determinations of \citet{shankar}  and \citet{Lauer} (at $z=0$), and those of \citet{merloni} and \citet{schulze} (at $z>0$).}
\label{MF}
\end{figure*}

\begin{figure*}[!tp]
\centering
\includegraphics[width=0.9\textwidth]{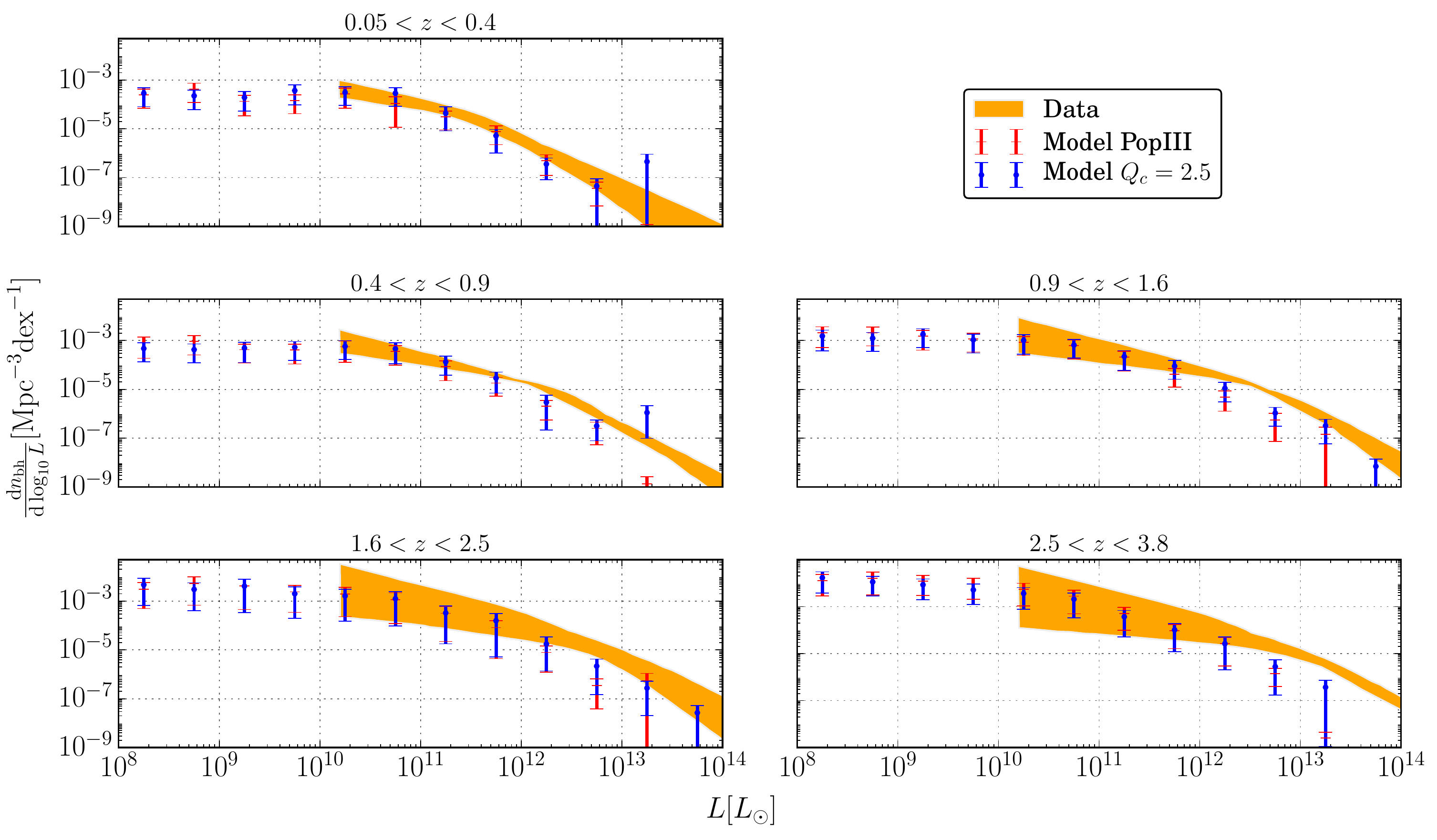}
\caption{Predictions of our model (with the same seed scenarios as in Fig.~\ref{MF}) for the bolometric AGN luminosity function, compared with observational determinations -- \citet{hopkins}, \citet{lacy}, \citet{lafranca}, and 
\citet{aird}, the last only considered at $z>3$ --, whose envelope is shown by a shaded orange area.}
\label{LF}
\end{figure*}

We compare the black hole mass function predicted by our semi-analytic galaxy formation model with the observational determinations of \citet{shankar}  and \citet{Lauer} (at $z=0$), and with those of \citet{merloni} and \citet{schulze} (at $z>0$) in Fig.~\ref{MF}. The model's predictions are shown as red bars or blue dots, the first referring to a scenario in which black holes form from low-mass seeds of a few hundred $M_\odot$ \citep[e.g., the remnants of Pop III stars;][]{MadauRees2001}, and the second representing a model in which black holes descend from ``heavy'' ($\sim 10^5 M_\odot$) seeds arising, for example, from instabilities of protogalactic disks. In more detail, for the latter case we use the model of \citet{Volonteri2008}, setting
their critical Toomre parameter, which regulates the onset of the instability, to their preferred value $Q_c=2.5$.) The error bars of the model's points are Poissonian.
As can be seen, the agreement with the data is rather good, especially in the mass range relevant for our purposes ($M_{\rm bh}< 10^8 M_\odot$).
As a further test, we  also compared the predictions of our model for the AGN (bolometric) luminosity function with the observations of \citet{hopkins}, \citet{lacy}, \citet{lafranca}, and 
\citet{aird}, whose envelope we show in Fig.~\ref{LF} as a shaded orange area. We note that we only consider the luminosity function of \citet{aird} at $z<3$ as it may be underestimated at larger redshifts \citep{kal}.

\end{document}